\documentclass[aps,pra,twocolumn,superscriptaddress,10pt,showpacs]{revtex4-1}

\usepackage{amsmath,amssymb,bbm}
\usepackage{graphicx}
\usepackage{dsfont}
\usepackage[T1]{fontenc}

\newcommand{\D}{\mathrm{d}}
\newcommand{\I}{\mathrm{i}}
\newcommand{\ds}{\displaystyle}

\newcommand{\Exp}[1]{\mathrm{e}^{\mbox{\footnotesize$#1$}}}
\newcommand{\power}[1]{^{\mbox{\footnotesize$#1$}}}
\newcommand{\rewop}[1]{_{\mbox{\footnotesize$#1$}}}
\newcommand{\tr}[1]{\mathrm{tr}{\left\{#1\right\}}}
\newcommand{\expect}[1]{{\left\langle{#1}\right\rangle}}
\newcommand{\cR}{\mathcal{R}}
\newcommand{\cI}{\mathcal{I}}
\newcommand{\cC}{\mathcal{C}}
\newcommand{\cov}{\mathrm{cov}}
\newcommand{\pML}{\widehat{p}^{\ }_{\textsc{ml}}}
\newcommand{\FML}{\widehat{F}^{\ }_{\textsc{ml}}}
\newcommand{\TML}{\widehat{T}^{\ }_{\textsc{ml}}}
\newcommand{\FBM}{\widehat{F}^{\ }_{\textsc{bm}}}
\newcommand{\rhoML}{\widehat{\rho}^{\ }_{\textsc{ml}}}
\newcommand{\rhoBM}{\widehat{\rho}^{\ }_{\textsc{bm}}}
\newcommand{\wref}{w_{\mathrm{r}}}
\newcommand{\pref}{P_{\mathrm{r},0}}
\newcommand{\prefD}{P_{\mathrm{r},D}}
\newcommand{\qref}{W_{\mathrm{r},0}}
\newcommand{\qw}{\widetilde{W}_{\mathrm{r},0}}
\newcommand{\qrefD}{W_{\mathrm{r},D}}
\newcommand{\ket}[1]{{\left|{#1}\right\rangle}}
\newcommand{\bra}[1]{{\left\langle{#1}\right|}}
\newcommand{\ketbra}[1]{\ket{#1}\bra{#1}}
\newcommand{\up}{\uparrow}
\newcommand{\dn}{\downarrow}

\newcommand{\chsh}{\theta}
\newcommand{\chshopt}{\theta_{\mathrm{opt}}}
\newcommand{\CHSH}{\Theta}
\newcommand{\CHSHopt}{\Theta_{\mathrm{opt}}}
\newcommand{\lamcr}{\lambda_{\mathrm{crit}}}

\newcommand{\tav}{t_{\mathrm{av}}}
\newcommand{\tmin}{t_{\mathrm{min}}}

\renewcommand{\eqref}[1]{(\ref{eq:#1})}
\newcommand{\figref}[1]{\ref{fig:#1}}

\newcommand{\STEP}[3][0.5\baselineskip]{\par\vspace*{#1}\par\noindent %
\textbf{#2}\ {#3}}

\DeclareMathAlphabet{\vecfont}{OT1}{cmr}{bx}{it}
\renewcommand{\vec}[1]{\vecfont{#1}}
\newcommand{\bfsym}[1]{\boldsymbol{#1}} 
\DeclareMathAlphabet{\dyadfont}{OT1}{cmss}{bx}{n}
\newcommand{\dyadic}[1]{\dyadfont{#1}}
\newcommand{\identity}{\dyadic{1}}

\begin{document}

\title{Optimal error intervals for properties of the quantum state}

\date[]{Posted on the arXiv on \today}  

\author{Xikun Li}
\affiliation{Centre for Quantum Technologies, National University of Singapore, %
3 Science Drive 2, Singapore 117543, Singapore}

\author{Jiangwei Shang}
\altaffiliation[Now at\ ]{Naturwissenschaftlich-Technische Fakult\"at, %
       Universit\"at Siegen, Walter-Flex-Stra\ss{}e 3, 57068 Siegen, Germany}
\email{corresponding email: jiangwei.shang@quantumlah.org}
\affiliation{Centre for Quantum Technologies, National University of Singapore, %
3 Science Drive 2, Singapore 117543, Singapore}

\author{Hui Khoon Ng}
\affiliation{Centre for Quantum Technologies, National University of Singapore, %
3 Science Drive 2, Singapore 117543, Singapore}
\affiliation{Yale-NUS College, 16 College Avenue West, Singapore 138527, Singapore}
\affiliation{MajuLab, CNRS-UNS-NUS-NTU International Joint Research Unit, %
UMI 3654, Singapore}

\author{Berthold-Georg Englert}
\affiliation{Centre for Quantum Technologies, National University of Singapore, %
3 Science Drive 2, Singapore 117543, Singapore}
\affiliation{MajuLab, CNRS-UNS-NUS-NTU International Joint Research Unit, %
UMI 3654, Singapore}
\affiliation{Department of Physics, National University of Singapore, %
2 Science Drive 3, Singapore 117542, Singapore}

\begin{abstract}
Quantum state estimation aims at determining the quantum state from observed
data.
Estimating the full state can require considerable efforts, but one is often
only interested in a few properties of the state, such as the fidelity with a
target state, or the degree of correlation for a specified bipartite
structure.
Rather than first estimating the state, one can, and should, estimate those
quantities of interest directly from the data.
We propose the use of optimal error intervals as a meaningful way of stating
the accuracy of the estimated property values.
Optimal error intervals are analogs of the optimal error regions for state
estimation [New J.~Phys.~\textbf{15}, 123026 (2013)].
They are optimal in two ways: They have the largest likelihood for the
observed data and the pre-chosen size, and are the smallest for the pre-chosen
probability of containing the true value.
As in the state situation, such optimal error intervals admit a simple
description in terms of the marginal likelihood for the data for the
properties of interest.
Here, we present the concept and construction of optimal error intervals,
report on an iterative algorithm for reliable computation of the marginal
likelihood (a quantity difficult to calculate reliably), explain how plausible
intervals --- a notion of evidence provided by the data --- are related to our
optimal error intervals, and illustrate our methods with single-qubit and
two-qubit examples.
\end{abstract}


\maketitle

\section{Introduction}\label{sec:Intro}
Quantum state estimation (QSE) --- the methods, procedures, and algorithms by
which one converts tomographic experimental data into an educated guess about the
state of the quantum system under investigation \cite{LNP649} --- provides
just that: an estimate of the \emph{state}.
For high-dimensional systems, such a state estimate can be hard to come by.
But one is often not even interested in all the details the state conveys and
rather cares only about the values of a few functions of the state.
For example, when a source is supposed to emit quantum systems in a specified
target state, the fidelity between the actual state and this target could be
the one figure of merit we want to know.
Then, a direct estimate of the few properties of interest,
without first estimating the quantum state, is more practical and more
immediately useful.

The full state estimate may not even be available in the first place, if only
measurements pertinent to the quantities of interest are made instead of a
tomographically complete set, the latter involving a
forbidding number of measurement settings in high dimensions.
Furthermore, even if we have a good estimate for the quantum state, the values
of the few properties of interest computed from this state may not be, and
often are not, the best guess for those properties (see an illustration of
this point in Sec.~\ref{sec:SCPR}).
Therefore, we need to supplement QSE with SPE --- state-property estimation,
that is: methods, procedures, and algorithms by which one  directly arrives
at an educated guess for the few properties of interest.

Several schemes have been proposed for determining particular
properties of the quantum state.
These are prescriptions for the measurement scheme, and/or estimation
procedure from the collected data.
For example, there are schemes for measuring the traces of powers of the
statistical operator, and then perform separability tests with the numbers thus
found \cite{Ekert+5:02,Horodecki+1:02,Bovino+5:05}.
Alternatively, one could use likelihood ratios for an educated guess whether
the state is separable or entangled \cite{BK+2:10}.
Other schemes are tailored for measuring the fidelity with particular target
states \cite{Somma+2:06,Guhne+3:07,Flammia+1:11},
yet another can be used for estimating the concurrence \cite{Walborn+4:06}.
Schemes for measuring other properties of the quantum state can be found by
Paris's method \cite{Paris:09}.

Many of these schemes are property specific, involving sometimes ad-hoc
estimation procedures well-suited for only those properties of interest.
Here, in full analogy to the \emph{state} error regions of
Ref.~\cite{Shang+4:13} for QSE, we describe general-purpose optimal error
intervals for SPE, from measurement data obtained from generic tomographic
measurements or property-specific schemes like those mentioned above.
Following the maximum-likelihood philosophy for statistical inference, these
error intervals give precise ``error bars'' around the maximum-likelihood
(point) estimator for the properties in question consistent with the data.
According to the Bayesian philosophy, they are intervals with a precise
probability (credibility) of containing the true property values.
As is the case for QSE error regions, these SPE error intervals are
optimal in two ways.
First, they have the largest likelihood for the data among all the intervals
of the same size.
Second, they are smallest among all regions of the same credibility.
Here, the natural notion of the size of an interval is its prior content,
i.e., our belief in the interval's importance before any data are taken; the
credibility of an interval is its posterior --- after taking the data into
account --- content.

We will focus on the situation in which a single property of the state is of
interest.
This is already sufficient for illustration, but is not a restriction of our
methods.
(Note: If there are several properties of interest and a consistent set of
values is needed, they should be estimated jointly, not one-by-one, to ensure
that constraints are correctly taken into account.)
The optimal error interval is a range of values for this property that answers
the question: Given the observed data, how well do we know the value of the
property?
This question is well answered by the above-mentioned generalization of the
maximum-likelihood point estimator to an interval of most-likely values, as
well as the dual Bayesian picture of intervals of specified credibility.
Our error interval is in contrast to other work \cite{Faist+1:16}
based on the frequentists' concept of
confidence regions/intervals, which answer a different question pertaining to
all possible data that could have been observed
but is not the right concept for drawing inference from the actual
data acquired in a single run (see Appendix \ref{sec:appCC}).

As we will see below, the concepts and strategies of the optimal error regions
for QSE \cite{Shang+4:13,Shang+4:15,Seah+4:15} carry over naturally to this
SPE situation.
However, additional methods are needed for the specific computational tasks of
SPE.
In particular, there is the technical challenge of computing the
property-specific likelihood:
In QSE, the likelihood for the data as a function over the state space is
straightforward to compute; in SPE, the relevant likelihood is the
property-specific \emph{marginal likelihood}, which requires an integration of
the usual (state) likelihood over the ``nuisance parameters'' that are not of
interest.
This can be difficult to compute even in classical statistics \cite{Bos:02}.
Here, we offer an iterative algorithm that allows for reliable estimation of
this marginal likelihood.

In addition, we point out the connection between our optimal error intervals
and \emph{plausible intervals}, an elegant notion of evidence for property
values supported by the observed data \cite{Evans:15}.
Plausible intervals offer a complementary understanding of our error
intervals:
Plausibility identifies a unique error interval that contains all values for
which the data are in favor of, with an associated critical credibility
value.

Here is a brief outline of the paper.
We set the stage in Sec.~\ref{sec:stage} where we introduce the
reconstruction space and review the notion of size and credibility of
a region in the reconstruction space.
Analogously, we identify the size and credibility of a range of property
values in Sec.~\ref{sec:SCPR}.
Then, the flexibility of choosing priors in the property-value space is
discussed in Sec.~\ref{sec:prior}.
With these tools at hand, we formulate in Sec.~\ref{sec:PE-OEI}
the point estimators as well as the optimal error intervals for SPE.
Section~\ref{sec:evidence} explains the connection to plausible regions and
intervals.
Section~\ref{sec:MCint} gives an efficient numerical algorithm that
solves the high-dimensional integrals for the size and credibility.
We illustrate the matter by simulated single-qubit and two-qubit
experiments in Secs.~\ref{sec:1qubit} and \ref{sec:2qubit},
and close with a summary.

Additional material is contained in several appendixes: The fundamental
differences between Bayesian credible intervals and the confidence intervals
of frequentism are the subject matter of Appendix \ref{sec:appCC}.
Appendixes \ref{sec:appA} and \ref{sec:appB} deals with the limiting power
laws of the prior-content functions that are studied numerically in
Sec.~\ref{sec:2qubit}.
For ease of reference, a list of the various prior densities is given in
Appendix \ref{sec:appC} and a list of the acronyms in Appendix \ref{sec:appD}.

\section{Setting the stage}\label{sec:stage}
As in Refs.~\cite{Shang+4:13,Shang+4:15,Seah+4:15}, we regard the probabilities
$p=(p_1,p_2,\dots,p_K)$ of a measurement with $K$ outcomes as the
basic parameters of the quantum state $\rho$.
The Born rule
\begin{equation}\label{eq:2-0a}
  p_k=\tr{\Pi_k\rho}=\expect{\Pi_k}
\end{equation}
states that the $k$th probability $p_k$ is the expectation value of the $k$th
probability operator $\Pi_k$ in state $\rho$.
Together, the $K$ probability operators constitute a
probability-operator measurement (POM),
\begin{equation}\label{eq:2-0b}
  \Pi_k\geq0\,,\quad\sum_{k=1}^K\Pi_k=\identity\,,
\end{equation}
where $\identity$ is the identity operator.

The POM is fully tomographic if we can infer a unique state $\rho$ when the
values of all $p_k$s are known.
If the measurement provides partial rather than full tomography, we choose a
suitable set of statistical operators from the state space, such that the
mapping ${p\leftrightarrow\rho(p)}$ is one-to-one;
this set is the reconstruction space $\cR_0$.
While there is no unique or best choice for the ``suitable set'' that makes
up $\cR_0$, the intended use for the state, once estimated, may
provide additional criteria for choosing the reconstruction space.
As far as QSE and SPE are concerned, however, the particulars of the mapping
$p\to\rho(p)$ do not  matter at all.
Yet, that there is such a mapping,
permits viewing a region $\cR$ in $\cR_0$ also as a
region in the probability space, and we use the same symbols in both cases
whenever the context is clear.
Note, however, that while the probability space --- in which the numerical
work is done --- is always convex, the reconstruction space of states may
or may not be.
Examples for that can be found in \cite{Rehacek+6:15} where various aspects of
the mapping ${p\leftrightarrow\rho(p)}$ are discussed in the context of
measuring pairwise complementary observables.

The parameterization of the reconstruction space in terms of the probabilities
gives us
\begin{equation}\label{eq:2-1}
  (\D \rho)=(\D p)\, w_0(p)
\end{equation}
for the volume element $\equiv$ prior element in $\cR_0$, where
\begin{equation}\label{eq:2-2}
  (\D p)=\D p_1\D p_2\ldots \D p_K \, w_{\mathrm{cstr}}(p)\,,
\end{equation}
is the volume element in the probability space.
The factor $w_{\mathrm{cstr}}(p)$ accounts for all the constraints that the
probabilities must obey, among them the constraints that follow from the
positivity of $\rho(p)$ in conjunction with the quantum-mechanical Born
rule.
Other than the mapping ${p\leftrightarrow\rho(p)}$, this is the \emph{only}
place where quantum physics is present in the formalism of QSE and SPE.
Yet, the quantum constraints in $w_{\mathrm{cstr}}(p)$ are the defining
feature that distinguishes quantum state estimation from non-quantum state
estimation.

Probabilities $p$ that obey the constraints are called ``physical'' or
``permissible''. $w_{\mathrm{cstr}}$ vanishes on the unphysical $p$s and is
generally a product of step functions and delta functions.
The factor $w_0(p)$ in Eq.~\eqref{2-1} is the prior density of our choice;
it reflects what we know about the quantum system before the data are taken.
Usually, the prior density $w_0(p)$ gives positive weight to the
finite neighborhoods of all states in $\cR_0$;
criteria for choosing the prior are reviewed in appendix A of
Ref.~\cite{Shang+4:13} --- ``use common sense'' is a guiding principle.
Although not really necessary, we shall assume that $w_0(p)$ and
$w_{\mathrm{cstr}}(p)$ are normalized,
\begin{equation}\label{eq:2-3}
  \int(\D p)=1\quad\mbox{and}\quad\int_{\cR_0}(\D\rho)=1\,,
\end{equation}
so that we do not need to exhibit normalizing factors in what follows.
Then, the size of a region ${\cR\subseteq\cR_0}$, that is: its prior content, is
\begin{equation}\label{eq:2-4}
  S_{\cR}=\int_{\cR}(\D\rho) =\int_\cR(\D p)\,w_0(p)\leq1\,,
\end{equation}
with equality only for ${\cR=\cR_0}$. 
This identification of size and prior content is natural 
in the context of state estimation; see \cite{Shang+4:13} for a
discussion of this issue.
While other contexts may very well have their own natural notions of size, 
such other contexts do not concern us here.

After measuring a total number of ${N=\sum_{k=1}^K n_k}$ copies of the quantum
system and observing the $k$th outcome $n_k$ times, the data $D$ are the
recorded sequence of outcomes (``detector clicks'').
The probability of obtaining $D$ is the point likelihood
\begin{equation}\label{eq:2-5}
  L\left(D|p\right)=p_1^{n_1}p_2^{n_2}\cdots p_K^{n_K}\,.
\end{equation}
In accordance with Sec.~2.3 in Ref.~\cite{Shang+4:13}, then, the joint
probability of finding $\rho(p)$ in the region $\cR$ and obtaining data $D$ is
\begin{eqnarray}\label{eq:2-6}
  \mathrm{Pr}\bigl(D\wedge \{\rho\in\cR\}\bigr)
 &=&\int_{\cR}(\D p)\,w_0(p)\,L(D|p)
\nonumber\\&=&L(D|\cR)S_{\cR}=C_{\cR}(D)L(D)\,,
\end{eqnarray}
with (i) the region likelihood $L(D|\cR)$,
(ii) the credibility --- the posterior content --- $C_{\cR}(D)$ of the region,
\begin{equation}\label{eq:2-6a}
C_\cR(D)=\frac{1}{L(D)}\int_\cR(\D p)\,w_0(p)\,L(D|p)\,,
\end{equation}
and (iii) the prior likelihood for the data
\begin{equation}\label{eq:2-7}
  L(D)= \int_{\cR_0}(\D p)\,w_0(p)\,L(D|p)\,.
\end{equation}

\section{Size and credibility of a range of property values}\label{sec:SCPR}
We wish to estimate a particular property, specified as a function $f(p)$ of
the probabilities, with values between $0$ and $1$,
\begin{equation}\label{eq:3-1}
  0\leq f(p)\leq1\,;
\end{equation}
the restriction to this convenient range can be easily lifted, of course.
Usually, there is at first a function $\tilde{f}(\rho)$ of the state $\rho$,
and ${f(p)=\tilde{f}\bigl(\rho(p)\bigr)}$ is the implied function of $p$.
We take for granted that the value of
$\tilde{f}(\rho)$ can be found without requiring information that
is not contained in the probabilities $p$.
Otherwise, we need to restrict $\tilde{f}(\rho)$ to $\rho$s in $\cR_0$.

\begin{figure}
\centerline{\includegraphics[width=0.95\columnwidth,clip=]{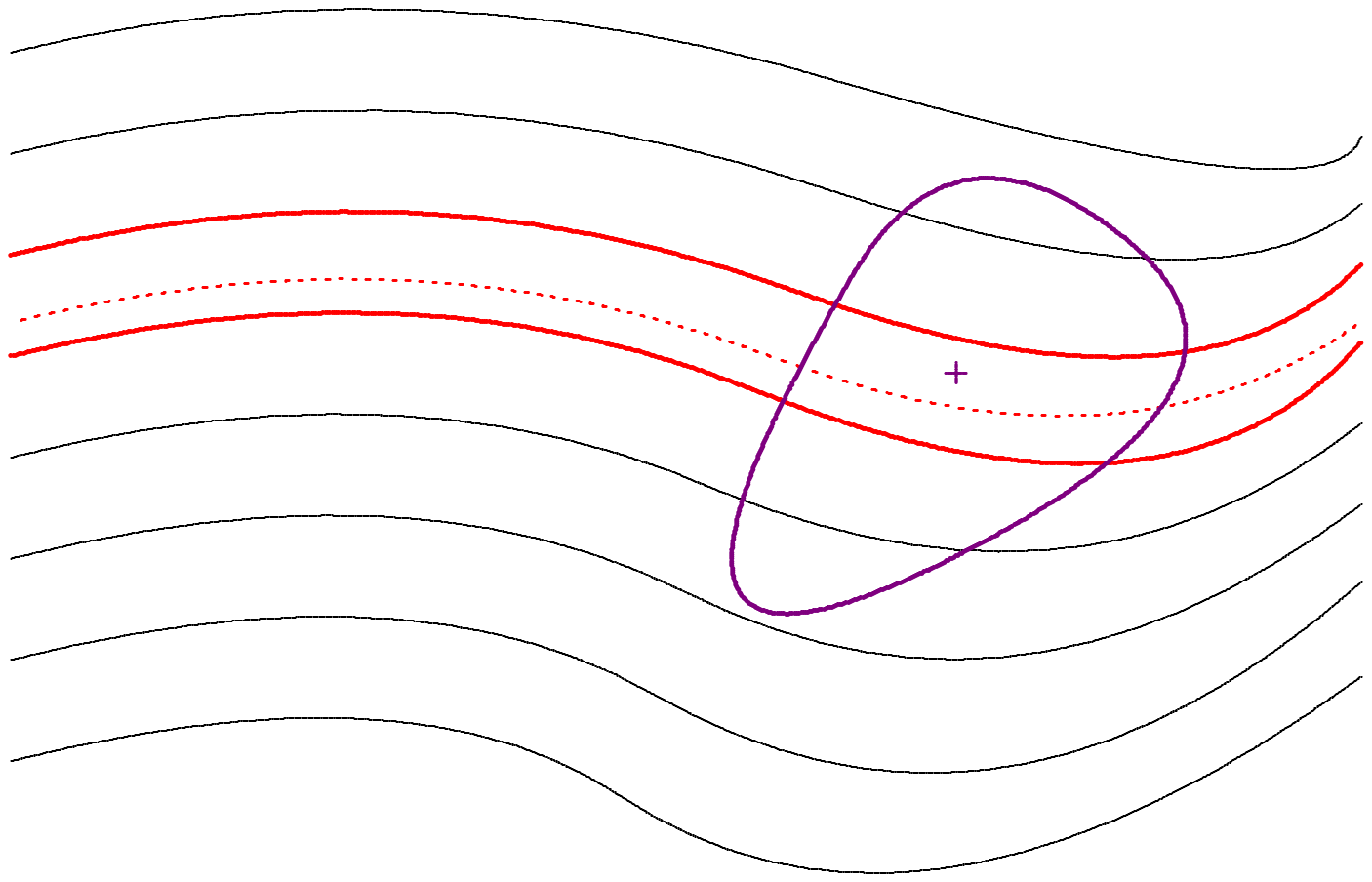}}
\caption{\label{fig:regions}
Schematic sketch of a sector in the probability
space or the reconstruction space.
The wave-like lines indicate iso-$F$ hypersurfaces; any two lines mark the
boundaries of an $F$ interval, a region specified by a range of $F$ values.
The thicker red lines mark the borders of a smallest credible interval
(SCI).
The dashed red line inside the SCI indicates the hypersurface of the
maximum-likelihood estimator $\FML$.
The purple cross marks the maximum-likelihood estimator $\rhoML$ of the
quantum
state, with the closed purple curve marking the boundary of the
smallest credible region (SCR) with the same credibility as the SCI.
The first equation in (\ref{eq:5-5}) states that the purple cross is
usually not on the dashed red line, as the plot shows.
Note that the SCR contains $F$ values from a larger range than the SCI;
see also Fig.~\figref{DSPEvsISPE}.}
\end{figure}

By convention, we use lower-case letters for the functions on the probability
space and upper-case letters for the function values.
The generic pair is $f(p),F$ here; we will meet the pairs $\phi(p),\Phi$ and
$\gamma(p),\Gamma$ in Sec.~\ref{sec:1qubit}, and the pairs $\chsh(p),\CHSH$
and  $\chshopt(p),\CHSHopt$ in Sec.~\ref{sec:2qubit}.

A given $f(p)$ value --- $F=f(p)$, say --- identifies hypersurfaces in the
probability space and the reconstruction space, and an interval
${F_1\leq f(p)\leq F_2}$ corresponds to a region; see Fig.~\figref{regions}.
Such a region has size
\begin{eqnarray}\label{eq:3-2}
  &&\int_{\cR_0}(\D\rho)\,\Bigl[\eta\bigl(F_2-\tilde{f}(\rho)\bigr)
                           -\eta\bigl(F_1-\tilde{f}(\rho)\bigr)\Bigr]
\nonumber\\&=&
\int(\D p)\,w_0(p)\Bigl[\eta\bigl(F_2-f(p)\bigr)
                           -\eta\bigl(F_1-f(p)\bigr)\Bigr]
\nonumber\\&=&
\int(\D p)\,w_0(p)\int_{F_1}^{F_2}\D F\,\delta\bigl(F-f(p)\bigr)
\end{eqnarray}
and credibility
\begin{eqnarray}\label{eq:3-3}
  &&\frac{1}{L(D)}\int_{\cR_0}(\D\rho)\,L(D|p)
                  \Bigl[\eta\bigl(F_2-\tilde{f}(\rho)\bigr)
                           -\eta\bigl(F_1-\tilde{f}(\rho)\bigr)\Bigr]
\nonumber\\&=&
\frac{1}{L(D)}\int(\D p)\,w_0(p)L(D|p)
           \int_{F_1}^{F_2}\D F\,\delta\bigl(F-f(p)\bigr)\,,
\end{eqnarray}
where $\eta(\,)$ is Heaviside's unit step function and $\delta(\,)$ is
Dirac's delta function.
For an infinitesimal slice, ${F\leq f(p)\leq F+\D F}$, the size \eqref{3-2}
identifies the prior element ${\D F\,W_0(F)}$ in $F$,
\begin{equation}\label{eq:3-4}
  \D F\,W_0(F)=\int(\D p)\,w_0(p)\,\D F\,\delta\bigl(F-f(p)\bigr)\,,
\end{equation}
and the credibility \eqref{3-3} tells us the likelihood $L(D|F)$ of the data
for given property value $F$,
\begin{eqnarray}\label{eq:3-5}
   &&\frac{1}{L(D)}\ \D F\,W_0(F)L(D|F)
   \nonumber\\&=&
\frac{1}{L(D)}\int(\D p)\,w_0(p)L(D|p)\,\D F\,\delta\bigl(F-f(p)\bigr)\,.
\end{eqnarray}
Of course, Eqs.~\eqref{3-4} and \eqref{3-5} are just the statements of
Eqs.~\eqref{2-4} and \eqref{2-6a} in the current context of infinitesimal
regions defined by an increment in $F$; it follows that $W_0(F)$ and $L(D|F)$
are positive everywhere, except possibly for a few isolated values of $F$.
To avoid any potential confusion with the likelihood $L(D|p)$ of
Eq.~\eqref{2-5}, we shall call $L(D|F)$ the $F$-likelihood.

In passing, we note that $L(D|F)$ can be viewed as the
marginal likelihood of $L(D|p)$ with respect to the probability density
$\delta\bigl(F-f(p)\bigr)/W_0(F)$ in $p$.
For the computation of $L(D|F)$, however, standard numerical methods for
marginal likelihoods, such as those compared by Bos \cite{Bos:02}, do not give
satisfactory results.
The bench marking conducted by Bos speaks for itself; in particular,
we note that none of those standard methods has a built-in accuracy check.
Therefore, we are using the algorithm described in Sec.~\ref{sec:MCint}.

In terms of $W_0(F)$ and $L(D|F)$, a finite interval of $F$ values, or the
union of such intervals, denoted by the symbol $\cI$, has the size
\begin{equation}\label{eq:3-6}
  S_{\cI}=\int_{\cI}\D F\,W_0(F)
\end{equation}
and the credibility
\begin{equation}\label{eq:3-7}
  C_{\cI}=\frac{1}{L(D)}\int_{\cI}\D F\,W_0(F)L(D|F)\,,
\end{equation}
where
\begin{equation}\label{eq:3-8}
  L(D)=\int_{\cI_0}\D F\,W_0(F)L(D|F)
\end{equation}
has the same value as the integral of Eq.~\eqref{2-7}.
$\cI_0$ denotes the whole range ${0\leq F\leq1}$ of property values, where we
have $S_{\cI_0}=C_{\cI_0}=1$.

Note that the $F$-likelihood $L(D|F)$ is the natural derivative of the
interval likelihood, the conditional probability
\begin{eqnarray}\label{eq:3-9}
  L(D|\cI)&=&\frac{\mathrm{Pr}\bigl(D\wedge \{F\in\cI\}\bigr)}
                  {\mathrm{Pr}(F\in\cI)}\\
  &=&\frac{1}{S_\cI}\int(\D p)\,w_0(p)\,L(D|p)\int_\cI\D F\,\delta(F-f(p))\,.
\nonumber
\end{eqnarray}
If we now define the $F$-likelihood by the requirement
\begin{equation}\label{eq:3-10}
L(D|\cI)=\frac{1}{S_\cI}\int_\cI \D F\,W_0(F)\,L(D|F)\,,
\end{equation}
we recover the expression for $L(D|F)$ in Eq.~\eqref{3-5}.

\section{Free choice of prior}\label{sec:prior}
The prior density $W_0(F)$ and the $F$-likelihood $L(D|F)$ have an implicit
dependence on the prior density $w_0(p)$ in probability space, and it may seem
that we cannot choose $W_0(F)$ as we like, nor would the $F$-likelihood
be independent of  the prior for $F$.
This is only apparently so:
As usual, the likelihood does not depend on the prior.

When we restrict the prior density $w_0(p)$ to the hypersurface where
${f(p)=F}$,
\begin{equation}\label{eq:4-0a}
  w_0(p)\Bigr|_{f(p)=F}=W_0(F)u_F(p)\,,
\end{equation}
we exhibit the implied prior density $u_F(p)$ that tells us the relative
weights of $p$s within the iso-$F$ hypersurface.
As a consequence of the normalization of $w_0(p)$ and $W_0(F)$,
\begin{equation}\label{eq:4-0b}
  \int(\D p)\,w_0(p)=1\,,\qquad\int_0^1\D F\,W_0(F)=1\,,
\end{equation}
which are more explicit versions of ${S_{\cR_0}=1}$ and ${S_{\cI_0}=1}$,
$u_F(p)$ is also normalized,
\begin{equation}\label{eq:4-1}
  \int (\D p)\,u_F(p)\,\delta(F-f(p))=1\,.
\end{equation}
In a change of perspective, let us now regard $u_F(p)$ and $W_0(F)$ as
independently chosen prior densities for all iso-$F$ hypersurfaces and for
property $F$.
Since $F$ is the coordinate in $p$-space that is normal to the iso-$F$
hypersurfaces (see Fig.~\figref{regions}), these two prior densities
together define a prior density on the whole probability space,
\begin{equation}\label{eq:4-3}
    w_0(p)=W_0(f(p))\,u_{f(p)}(p)\,.
\end{equation}
The restriction to a particular value of $f(p)$ takes us back to Eq.~\eqref{4-0a},
as it should.

For a prior density of the form \eqref{4-3}, the $F$-likelihood
\begin{eqnarray}\label{eq:4-0c}
  L(D|F)&=&\frac{1}{W_0(F)}\int(\D p)\, w_0(p)L(D|p)\delta\bigl(F-f(p)\bigr)
\nonumber\\
        &=&\int(\D p)\, u_F(p)L(D|p)\delta\bigl(F-f(p)\bigr)
\end{eqnarray}
does not involve $W_0(F)$ and is solely determined by $u_F(p)$.
Therefore, different choices for $W_0(F)$ in Eq.~\eqref{4-3} do not result in
different $F$-likelihoods.
Put differently, if we begin with some reference prior density
$\wref(p)$, which yields the iso-$F$ prior density
\begin{equation}\label{eq:4-0e}
   u_F(p)=\frac{\ds\wref(p)\Bigr|_{f(p)=F}}
               {\ds\int(\D p')\,\wref(p')\,\delta\bigl(F-f(p')\bigr)}
\end{equation}
that we shall use throughout, then
\begin{equation}\label{eq:4-0d}
  w_0(p)=\frac{\wref(p)W_0(f(p))}
              {\ds\int(\D p')\,\wref(p')\,\delta\bigl(f(p)-f(p')\bigr)}
\end{equation}
is the corresponding prior density for the $W_0(F)$ of our liking.
Clearly, the normalization of $w_{\mathrm{r}}(p)$ is not important; more
generally yet, the replacement
\begin{equation}\label{eq:4-0f}
  \wref(p)\to\wref(p)g\bigl(f(p)\bigr)
\end{equation}
with an arbitrary function ${g(F)>0}$ has no effect on the right-hand sides of
Eqs.~\eqref{4-0e}, \eqref{4-0d}, as well as \eqref{4-5} below.
One can think of this replacement as modifying the prior density in $F$ that
derives from $\wref(p)$ upon proper normalization.

While the $F$-likelihood
\begin{equation}\label{eq:4-5}
  L(D|F)=\frac{\ds\int(\D p)\,\wref(p)\,\delta\bigl(F-f(p)\bigr)L(D|p)}
              {\ds\int(\D p)\,\wref(p)\,\delta\bigl(F-f(p)\bigr)}
\end{equation}
is the same for all $W_0(F)$s, it will
usually be different for different $u_F(p)$s and thus for different
$\wref(p)$s.
For sufficient data, however, $L(D|p)$ is so narrowly peaked in probability
space that it will be essentially vanishing outside a small region within the
iso-$F$ hypersurface, and then it is irrelevant which reference prior is used.
In other words, the data dominate rather than the priors unless the data are
too few.

Typically, we will have a natural choice of prior density $w_0(p)$ on the
probability space and accept the induced $W_0(F)$ and $u_F(p)$.
Nevertheless, the flexibility offered by Eq.~\eqref{4-0d} is useful.
We exploit it for the numerical procedure in Sec.~\ref{sec:MCint}.

In the examples below, we employ two different reference priors $\wref(p)$.
The first is the \emph{primitive prior},
\begin{equation}\label{eq:4-6}
  w_{\mathrm{primitive}}(p)= 1\,,
\end{equation}
so that the density is uniform in $p$ over the (physical) probability space.
The second is the \emph{Jeffreys prior} \cite{Jeffreys:46},
\begin{equation}\label{eq:4-7}
  w_{\mathrm{Jeffreys}}(p)\propto
   \frac1{\sqrt{p_{1}p_{2}\cdots p_{K}}}\,,
\end{equation}
which is a common choice of prior when no specific prior information is
available \cite{Kass+1:96}.
For ease of reference, there is a list of the various prior densities in
Appendix~\ref{sec:appC}.

In Sec.~\ref{sec:1qubit}, we use
$w_{\mathrm{primitive}}(p)$ and $w_{\mathrm{Jeffreys}}(p)$
for $w_0(p)$ and then work with the induced priors $W_0(F)$ of Eq.~\eqref{3-4}, as
this enables us to discuss the difference between direct and indirect
estimation in Sec.~\ref{sec:1qubitb}.
The natural choice of ${W_0(F)=1}$ will serve as the prior density in
Sec.~\ref{sec:2qubit}.

\section{Point estimators and optimal error intervals}\label{sec:PE-OEI}
The $F$-likelihood $L(D|F)$ is largest for the maximum-likelihood estimator
$\FML$,
\begin{equation}\label{eq:5-1}
  \max_F\{L(D|F)\}=L(D|\FML)\,.
\end{equation}
Another popular point estimator is the Bayesian mean estimator
\begin{equation}\label{eq:5-2}
  \FBM=\frac{1}{L(D)}\int_0^1\D F\,W_0(F)\,L(D|F)\,F\,.
\end{equation}
They are immediate analogs of the maximum-likelihood estimator $\rhoML$ for
the state,
\begin{equation}\label{eq:5-3}
  \rhoML=\rho(\pML)\quad\mbox{with}\quad\max_p\{L(D|p)\}=L(D|\pML)\,,
\end{equation}
and the Bayesian mean of the state,
\begin{equation}\label{eq:5-4}
  \rhoBM=\frac{1}{L(D)}\int(\D\rho)\,L(D|p)\,\rho\,.
\end{equation}
Usually, the value of $\tilde{f}(\rho)$ for one of these state estimators is
different from the corresponding estimator,
\begin{equation}\label{eq:5-5}
  \tilde{f}(\rhoML)\neq\FML\,,\quad\tilde{f}(\rhoBM)\neq\FBM\,,
\end{equation}
although the equal sign can hold for particular data $D$;
see Fig.~\figref{regions}.
As an exception, we note that $\tilde{f}\left(\rhoBM\right)=\FBM$
is always true if $f(p)$ is linear in $p$.

The observation of Eq.~\eqref{5-5} --- the
best guess for the property of interest may not, and often does
not, come from the best guess for the quantum state ---
deserves emphasis, although it is not a new insight.
For example, the issue is discussed in Ref.~\cite{Schwemmer+6:15} in the
context of confidence regions (see topic SM4 in the supplemental material).
We return to this in Sec.~\ref{sec:1qubitb}.

For reasons that are completely analogous to those for the optimal error
regions in Ref.~\cite{Shang+4:13}, the optimal error intervals for property
${F=\tilde{f}(\rho)=f(p)}$ are the bounded-likelihood intervals (BLIs)
specified by
\begin{equation}\label{eq:5-6}
  \cI_{\lambda}=\left\{F\bigm|L(D|F)\geq\lambda L(D|\FML)\right\}
\quad\mbox{with}\quad0\leq\lambda\leq1\,.
\end{equation}
While the set of $\cI_{\lambda}$s is fully specified by the
  $F$-likelihood $L(D|F)$ and is independent of the prior density $W_0(F)$,
  the size and credibility of a specific $\cI_{\lambda}$ do depend on the
  choice of $W_0(F)$.
The interval of largest $F$-likelihood for given size $s$ --- the
maximum-likelihood interval (MLI) --- is the BLI with
${\ds s=S_{\cI_\lambda}\equiv s_\lambda}$, and the interval of smallest size
for given credibility~$c$ --- the smallest credible interval (SCI) ---
is the BLI with ${\ds c=C_{\cI_{\lambda}}}\equiv c_{\lambda}$, where $S_{\cI_{\lambda}}$
and $C_{\cI_{\lambda}}$ are the size and credibility of Eqs.~\eqref{3-6} and
\eqref{3-7} evaluated for the interval $\cI_{\lambda}$.
We have ${\cI_\lambda\subseteq\cI_0}$, ${s_\lambda\leq s_0=1}$, and ${c_\lambda\leq
c_0=1}$ for ${\lambda\leq\lambda_0}$, with ${\lambda_0\geq0}$ given by
${\min_F\{L(D|F)\}=\lambda_0 L(D|\FML)}$.
As $\lambda$ increases from $\lambda_0$ to $1$, $s_\lambda$ and $c_\lambda$
decreases monotonically from $1$ to $0$.
Moreover, we have the link between $s_{\lambda}$ and $c_{\lambda}$,
\begin{equation}\label{eq:5-7}
 c_{\lambda}=\frac{\ds\lambda s_{\lambda}+\int_\lambda^1 \D \lambda'
 \,s_{\lambda'}} {\ds\int_0^1 \D \lambda' \,s_{\lambda'}}\,,
\end{equation}
exactly as that for the size and credibility of bounded-likelihood regions
(BLRs) for state estimation in Ref.~\cite{Shang+4:13}.
The normalizing integral of the size in the denominator has a particular
significance of its own, as is discussed in the next section.

As soon as the $F$-likelihood $L(D|F)$ is at hand, it is a simple matter
to find the MLIs and the SCIs.
Usually, we are most interested in the SCI for the desired credibility~$c$:
The actual value of $F$ is in this SCI with probability $c$.
Since all BLIs contain the maximum-likelihood estimator $\FML$, each BLI,
and thus each SCI, reports an error bar on $\FML$ in this precise sense.
In marked contrast, $\FBM$ plays no such distinguished role.

\section{Plausible regions and intervals}\label{sec:evidence}
The data provide evidence in favor of the $\rho$s in a region $\cR\subset\cR_0$ if we
would put a higher bet on $\cR$ after the data are recorded than before, that
is: if the credibility of $\cR$ is larger
than its size,
\begin{equation}\label{eq:5-8}
 C_{\cR}(D)=\int_\cR(\D p)\,w_0(p)\frac{L(D|p)}{L(D)}
> \int_\cR(\D p)\,w_0(p)=S_{\cR}\,.
\end{equation}
In view of Eq.~\eqref{2-6}, this is equivalent to requiring that the region
likelihood $L(D|\cR)$ exceeds $L(D)$, the likelihood for the data.

Upon considering an infinitesimal vicinity of a state $\rho\leftrightarrow p$,
we infer from Eq.~\eqref{5-8} that we have \emph{evidence in favor}
of $\rho(p)\in\cR_0$ if ${L(D|p)>L(D)}$, and we have \emph{evidence against}
$p$, and thus against $\rho(p)$, if $L(D|p)<L(D)$.
The ratio $L(D|p)/L(D)$, or any monotonic function of it, measures the
strength of the evidence \cite{Evans:15}.
It follows that the data provide strongest evidence for the maximum-likelihood
estimator.

Further, since ${c_{\lambda}>s_{\lambda}}$ for all BLRs,
there is evidence in favor of each BLR.
The larger BLRs, however, those for the lower likelihood thresholds set by
smaller $\lambda$ values, contain subregions against which the data give
evidence.

The $\rho(p)$s with evidence against them are not plausible guesses for the
actual quantum state.
We borrow Evans's terminology \cite{Evans:15} and call the set of all $\rho$s,
for which the data provide evidence in favor, the \emph{plausible region} ---
the largest region with evidence in favor of all subregions.
It is the SCR $\cR_{\lambda}$ for the \emph{critical value} of $\lambda$,
\begin{subequations}\label{eq:5-9}
\begin{eqnarray}\label{eq:5-9a}
  &&\lamcr\equiv\frac{L(D)}{L_{\mathrm{max}}(D)}=\int_0^1 \D \lambda
  \,s_{\lambda}
\\\label{eq:5-9b}
\mbox{with}\quad&&L_{\mathrm{max}}(D)=\max_p\bigl\{L(D|p)\bigr\}\,.
\end{eqnarray}
\end{subequations}
The equal sign in Eq.~\eqref{5-9a} is that of Eq.~(21) in Ref.~\cite{Shang+4:13}.
In a plot of $s_{\lambda}$ and $c_{\lambda}$, such as those in Figs.~4 and 5
of \cite{Shang+4:13} or in Figs.~\figref{Size&Credibility} and
\figref{BLI-SizeCred} below, we can identify $\lamcr$ as the $\lambda$ value
with the largest difference $c_{\lambda}-s_{\lambda}$.

This concept of the plausible region for QSE carries over to SPE, where we
have the \emph{plausible interval} composed of those $F$ values for which
$L(D|F)$ exceeds $L(D)$.
It is the SCI $\cI_{\lambda}$ for the critical $\lambda$ value,
\begin{subequations}\label{eq:5-10}
\begin{eqnarray}\label{eq:5-10a}
  &&\lamcr\equiv\frac{L(D)}{L_{\mathrm{max}}(D)}=\int_0^1 \D \lambda
  \,s_{\lambda}
\\\label{eq:5-10b}
\mbox{with}\quad&&L_{\mathrm{max}}(D)=\max_{F}\bigl\{L(D|F)\bigr\}\,,
\end{eqnarray}
\end{subequations}
where now $L_{\mathrm{max}}(D)$ and $s_{\lambda}$ refer to the $F$-likelihood
$L(D|F)$.
Usually, the values of $L_{\mathrm{max}}(D)$ in Eqs.~\eqref{5-9b} and
\eqref{5-10b} are different and, therefore, the critical $\lambda$ values are
different.

After measuring a sufficient number of copies of the quantum system ---
symbolically: ``${\,N\gg1\,}$'' --- one can invoke the central limit theorem
and approximate the $F$-likelihood by a gaussian with a width $\propto
N^{-1/2}$,
\begin{equation}\label{eq:5-11}
  N\gg1\,:\quad L(D|F)\simeq
   L_{\mathrm{max}}(D)\,\Exp{-\frac{N}{2\alpha^2}(F-\FML)^2}\,,
\end{equation}
where ${\alpha>0}$ is a scenario-dependent constant.
The weak $N$-dependence of $\alpha$ and $\FML$ is irrelevant here and will be
ignored.
Then, the critical $\lambda$ value is
\begin{equation}\label{eq:5-12}
   N\gg1\,:\quad \lamcr\simeq W_0(\FML)\alpha\sqrt{\frac{2\pi}{N}}\,,
\end{equation}
provided that $W_0(F)$ is smooth near $\FML$, which property we take for
granted.
Accordingly, the size and credibility of the plausible interval are
\begin{equation}\label{eq:5-13}
   N\gg1\,:\quad\left\{\begin{array}{r@{\;\simeq\;}l}\ds
       s_{\lamcr}^{\ }&\ds
        2\lamcr{\left(\frac{1}{\pi}\log\frac{1}{\lamcr}\right)}^{\frac{1}{2}}\,,
 \\[2.5ex] \ds c_{\lamcr}^{\ }
   &\ds \mathrm{erf}{\left({\left(\log\frac{1}{\lamcr}\right)}^{\frac{1}{2}}\right)}
     \end{array}\right.
\end{equation}
under these circumstances.
When focusing on the dominating $N$ dependence, we have
\begin{equation}\label{eq:5-14}
   N\gg1\,:\quad \lamcr\,,\ s_{\lamcr}^{\ }\,,\ 1-c_{\lamcr}^{\ }
\propto\frac{1}{\sqrt{N}}\,,
\end{equation}
which conveys an important message:
As more and more copies of the quantum system are measured, the plausible
interval is losing in size and gaining in credibility.

\section{Numerical procedures}\label{sec:MCint}
The size element of Eq.~\eqref{3-4}, the credibility element of
Eq.~\eqref{3-5}, and the $F$-likelihood of Eqs.~\eqref{4-0c} and \eqref{4-5},
introduced in Eq.~\eqref{3-5}, are the core
ingredients needed for the construction of error intervals for $F$.
The integrals involved  are usually high-dimensional and can only be
computed by Monte Carlo (MC) methods.
The expressions with the delta-function factors in their integrands are,
however, ill-suited for a MC integration.
Therefore, we consider the antiderivatives
\begin{equation}\label{eq:6-1}
  \pref(F)=\int(\D p)\,\wref(p)\,\eta\bigl(F-f(p)\bigr)
\end{equation}
and
\begin{equation}\label{eq:6-2}
  \prefD(F)=\frac{1}{L(D)}
         \int(\D p)\,\wref(p)L(D|p)\,\eta\bigl(F-f(p)\bigr)\,.
\end{equation}
These are the prior and posterior contents of the interval ${0\leq f(p)\leq F}$
for the reference prior with density $\wref(p)$.
The denominator in the $F$-likelihood of Eq.~\eqref{4-5} is the derivative of
$\pref(F)$ with respect to $F$, the numerator that of $L(D)\prefD(F)$.

Let us now focus on the denominator in Eq.~\eqref{4-5},
\begin{equation}\label{eq:6-3}
  \qref(F)=\frac{\partial}{\partial F}\pref(F)
        =\int(\D p)\,\wref(p)\,\delta\bigl(F-f(p)\bigr)\,.
\end{equation}
For the MC integration, we sample the probability space in accordance with
the prior $\wref(p)$ and due attention to $w_{\mathrm{cstr}}(p)$ of
Eq.~\eqref{2-2}, for which the methods described in Refs.~\cite{Shang+4:15} and
\cite{Seah+4:15} are suitable.
This gives us $\pref(F)$ together with fluctuations that
originate in the random sampling and the finite size of the sample;
for a sample with $N_{\mathrm{sample}}$ values of $p$, the
  expected mean-square error is
$\bigl[\pref(F)\bigl(1-\pref(F)\bigr)/N_{\mathrm{sample}}\bigr]^{1/2}$.
We cannot differentiate this numerical approximation of $\pref(F)$,
but we can fit a several-parameter function to the values produced by
the MC integration, and then differentiate this function and so arrive at
an approximation $\qw(F)$ for $\qref(F)$.

How can we judge the quality of this approximation?
For the prior density $w_0(p)$ in Eq.~\eqref{4-0d} with any chosen
$W_0(F)$ \cite{note6},
the antiderivative of the integral in Eq.~\eqref{3-4} yields
\begin{eqnarray}\label{eq:6-4}
  P_0(F)&=&\int(\D p)\,w_0(p)\,\eta\bigl(F-f(p)\bigr)\nonumber\\&=&
  \int(\D p)\,\frac{\wref(p)W_0\bigl(f(p)\bigr)}{\qref\bigl(f(p)\bigr)}
              \int_0^F\D F'\,\delta\bigl(F'-f(p)\bigr)\nonumber\\&=&
  \int_0^F\frac{\D F'\, W_0(F')}{\qref(F')}\int(\D p)\,\wref(p)\,
                                          \delta\bigl(F'-f(p)\bigr)\nonumber\\
&=&\int_0^F\D F'\, W_0(F')
\end{eqnarray}
upon recalling Eq.~\eqref{6-3}.
When the approximation
\begin{equation}\label{eq:6-4a}
  \widetilde{w}_0(p)=\frac{\wref(p)W_0\bigl(f(p)\bigr)}{\qw\bigl(f(p)\bigr)}
\end{equation}
is used instead, we find
\begin{eqnarray}\label{eq:6-5}
  \widetilde{P}_0(F)&=&\int(\D p)\,\widetilde{w}_0(p)\,\eta\bigl(F-f(p)\bigr)
\nonumber\\&=&\int_0^F\!\D F'\,\frac{\qref(F')}{\qw(F')}W_0(F')\,.
\end{eqnarray}
It follows that $\qw(F)$ approximates $\qref(F)$ well if
${\widetilde{P}_0(F)\simeq \int_0^F\D F'\,W_0(F')}$ is sufficiently accurate.
If it is not, an approximation $\widetilde{W}_0(F)$ for
$\ds\frac{\partial}{\partial F}\widetilde{P}_0(F)$ provides
$\qw(F)\Bigr|_{\mathrm{new}}=\qw(F)\widetilde{W}_0(F)/W_0(F)$,
which improves on the approximation $\qw(F)$.
It does not give us $\qref(F)$ exactly because the integral in Eq.~\eqref{6-5}
also requires a MC integration with its intrinsic noise.

Yet, we have here the essence of an iteration algorithm for successive
approximations of $\qref(F)$.
Since the $F$-likelihood $L(D|F)$ does not depend on the prior $W_0(F)$,
we can choose ${W_0(F)=1}$ so that ${P_0(F)=F}$ in Eq.~\eqref{6-4}, and the
$n$th iteration of the algorithm consists of these steps:
\STEP{S1}{For given $\qref^{(n)}(F)$, sample the probability space in
  accordance with the prior
  ${w^{(n)}_0(p)=\wref(p)/\qref^{(n)}\bigl(f(p)\bigr)}$.}
\STEP{S2}{Use this sample for a MC integration of
  \begin{displaymath}
    P^{(n)}_0(F)=\int(\D p)\,w^{(n)}_0(p)\,\eta\bigl(F-f(p)\bigr)\,.
  \end{displaymath}}
\STEP{S3}{Escape the loop if ${P^{(n)}_0(F)\simeq F}$ with the desired
  accuracy.}
\STEP{S4}{Fit a suitable several-parameter function to the MC values
  of   $P^{(n)}_0(F)$.}
\STEP{S5}{Differentiate this function to obtain
  $$W^{(n)}_0(F)\simeq \frac{\partial}{\partial F}P^{(n)}_0(F)\,;$$
  update ${n\to n+1}$ and $$\qref^{(n)}(F)\to
    \qref^{(n+1)}(F)=\qref^{(n)}(F)W^{(n)}_0(F)\,;$$ return to step S1.}
\par\bigskip\par\noindent%
The sampling in step S1 consumes most of the CPU time in each round of
iteration.
It is, therefore, economic to start with smaller samples and increase the
sample size as the approximation gets better.
Numerical codes for sampling by the Hamiltonian MC method described in
Ref.~\cite{Seah+4:15} are available at a website \cite{QSampling}, where one also
finds large ready-for-use samples for a variety of POMs and priors.

Similarly, we compute the numerator $\qrefD(F)$ in Eq.~\eqref{4-5}.
With the replacements ${\qref^{(n)}(F)\to \qrefD^{(n)}(F)}$
and ${\wref(p)\to \wref(p)L(D|p)}$,
the same iteration algorithm works.
Eventually, we get the $F$-likelihood,
\begin{equation}\label{eq:6-6}
  L(D|F)=\frac{\qrefD(F)}{\qref(F)}\,,
\end{equation}
and can then proceed to determine the BLIs of Sec.~\ref{sec:PE-OEI}.

In practice, it is not really necessary to iterate until $P^{(n)}_0(F)$
equals $F$ to a very high accuracy.
A few rounds of the iteration are usually enough for establishing a
$w^{(n)}_0(p)$ for which the induced prior density $\qref^{(n)}(F)$ is
reliable over the whole range from ${F=0}$ to ${F=1}$.
Then a fit to  $\prefD(F)$, obtained from a MC integration with a sample in
accordance with the posterior density $\propto w^{(n)}_0(p)L(D|p)$, provides
an equally reliable posterior density $\qrefD^{(n)}(F)$, and so gives us the
$F$-likelihood of Eq.~\eqref{6-6}.
Regarding the fitting of a several-parameter function in step S4, we note
that, usually, a truncated Fourier series of the form
\begin{eqnarray}\label{eq:6-7}
  P_0^{(n)}(F)&\simeq&F+a_1\sin(\pi F)+a_2\sin(2\pi F)\nonumber\\
&&\phantom{F}+a_3\sin(3\pi F)+\cdots\,,
\end{eqnarray}
with the amplitudes $a_1,a_2,a_3,\dots$ as the fitting parameters, is a good
choice, possibly modified such that known properties of
$P_0^{(n)}(F)$ are properly taken into account.
These matters are illustrated by the examples in Sec.~\ref{sec:2qubit};
see, in particular, Fig.~\figref{P0-Iteration}.

\section{Example: One qubit}\label{sec:1qubit}
%
\begin{figure*}[!t]\centering
  \includegraphics{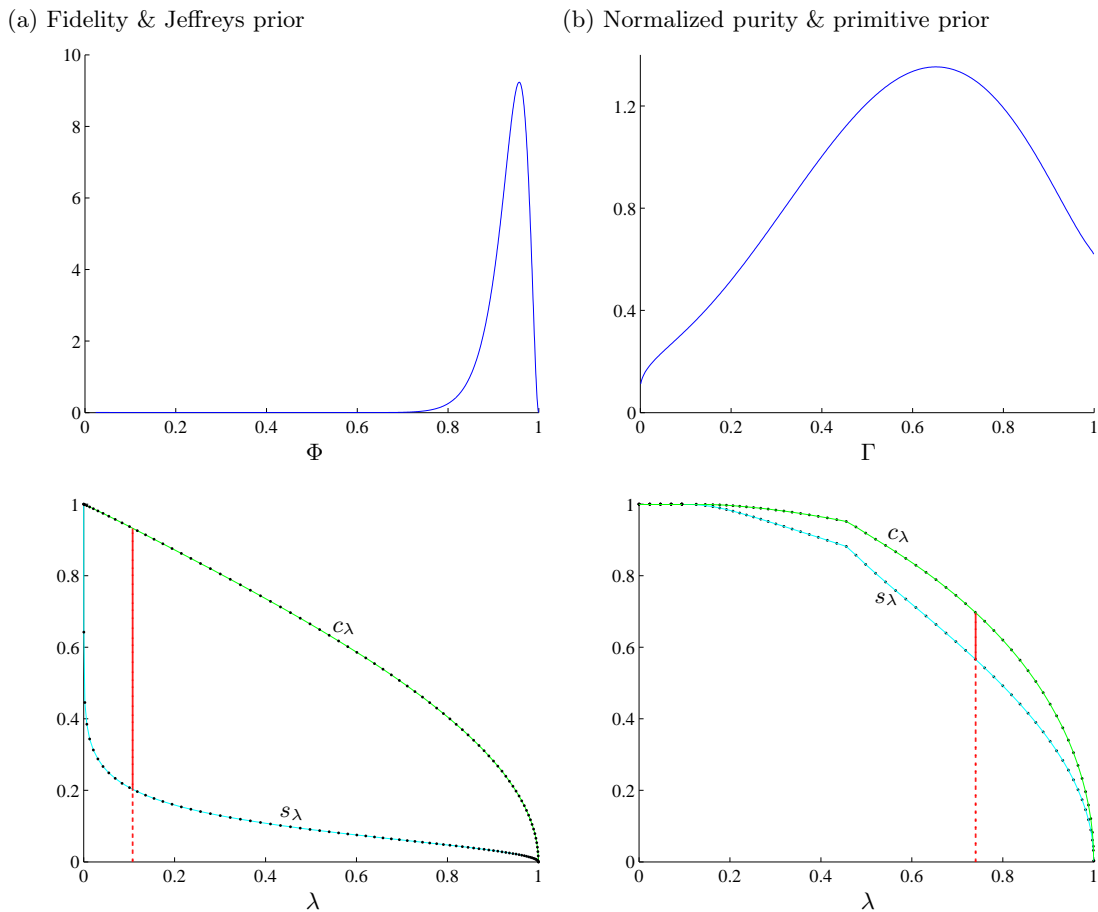}
  \caption{\label{fig:Size&Credibility}%
  Single-qubit fidelity (with respect to $|0\rangle$) and normalized
  purity for a simulated tetrahedron measurement of 36 copies.
  Top plots:
  The $\Phi$-likelihood $L(D|\Phi)$ and the $\Gamma$-likelihood $L(D|\Gamma)$
  for, respectively, the Jeffreys prior and the primitive prior on the
  probability space.
  Bottom plots:
  The size $s_\lambda$ (cyan curves) and the credibility $c_\lambda$
  (green curves) for the resulting BLIs as functions of $\lambda$.
  The black dots mark values obtained from the Hamiltonian Monte
  Carlo algorithm for evaluating the size and credibility integrals.
  The cyan lines are fitted to the $s_{\lambda}$ values using a Pad\'e
  approximant,
  while the green lines for $c_\lambda$ are obtained from the cyan lines
  with the aid of Eq.~\eqref{5-7}.
  The red vertical lines in the bottom plots mark the critical values of
  $\lambda$   at $\lamcr=0.1085$ and $\lamcr=0.7406$, respectively.}
\end{figure*}

As a first application, let us consider the single-qubit situation.
The state of a qubit can be written as
\begin{equation}\label{eq:7-4a}
\rho(\vec{r})=\frac{1}{2}(\identity+\vec{r}\cdot\bfsym{\sigma}),
\end{equation}
where $\bfsym{\sigma}=(\sigma_x,\sigma_y,\sigma_z)$ is the vector of Pauli
operators, and ${\vec{r}=(x,y,z)}$ is the Bloch vector with
${x=\langle\sigma_x\rangle}$, ${y=\langle\sigma_y\rangle}$, and
${z=\langle\sigma_z\rangle}$.
The tomographic measurement is taken to be the four-outcome tetrahedron
measurement of Ref.~\cite{Rehacek+1:04}, with outcome operators
\begin{equation}\label{eq:7-4}
  \Pi_k=\frac{1}{4}(\identity+\vec{a}_k\cdot\bfsym{\sigma})
     \quad \mathrm{with}\,\,k=1,2,3,4\,.
\end{equation}
Here, the four unit vectors $\vec{a}_k$ are chosen such that they are
respectively orthogonal to the four faces of a symmetric tetrahedron (hence
the name).
We orient them such that the probabilities
${p_k=\frac{1}{4}(1+\vec{r}\cdot\vec{a}_k)}$ for the four outcomes are
\begin{eqnarray}\label{eq:7-5}
  &&p_1=\frac{1}{4}(1-z)\,,\qquad
 p_2=\frac{1}{4}\!\left(1+\frac{\sqrt{8}}{3}y+\frac{1}{3}z\right),\nonumber\\
&&\begin{array}{@{}l@{}}p_3\\p_4\end{array}\Biggr\}
   =\frac{1}{4}\!\left(1\pm \sqrt{\frac{2}{3}}x
                     -\frac{\sqrt 2}{3}y+\frac{1}{3}z\right).
\end{eqnarray}
The tetrahedron measurement is tomographically complete for the qubit and
so allows the full reconstruction of the state, which we accomplish
with the aid of
\begin{equation}\label{eq:7-5'}
  \vec{r}=3\sum_{k=1}^4p_k\vec{a}_k\,.
\end{equation}
This tomographic completeness is useful for our discussion, since it permits
both the estimation of a property of interest directly from the $p_k$s, as
well as estimating that property by first estimating the density operator
$\rho$; see Sec.~\ref{sec:1qubitb}.

\subsection{SCIs for fidelity and purity}\label{sec:1qubita}
%
\begin{figure}[!t]\centering
  \includegraphics{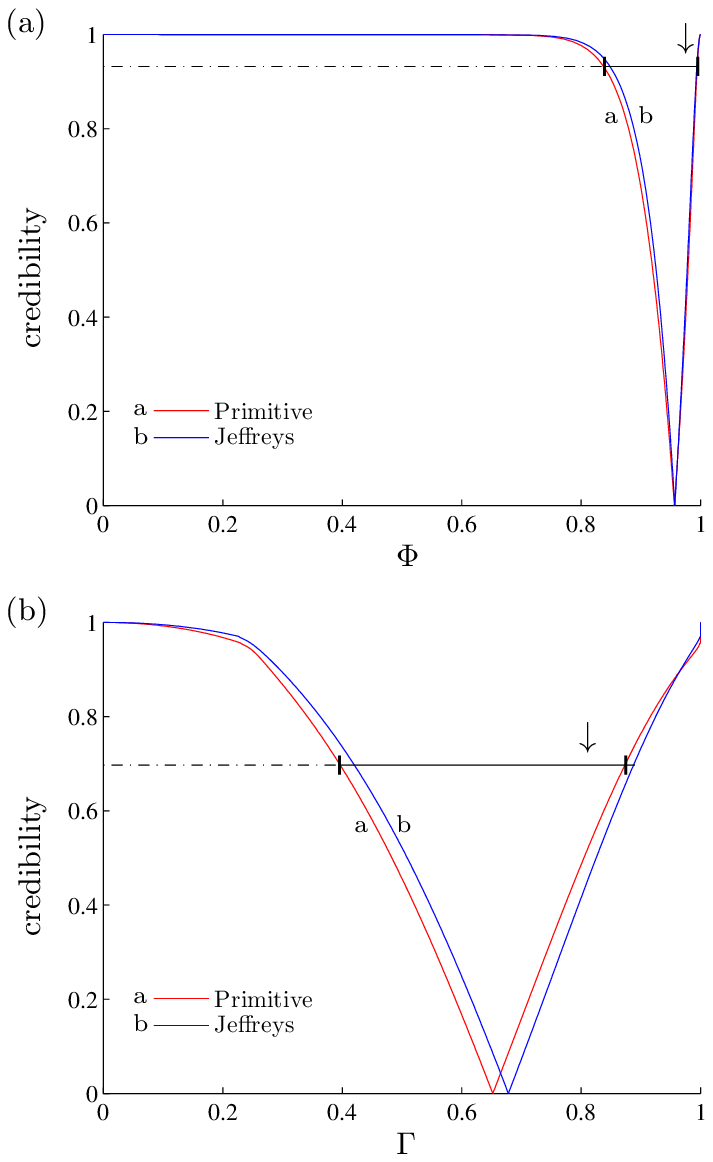}
  \caption{\label{fig:qubitSCIs}%
  Optimal error intervals
  for (a) fidelity $\Phi$, and
  (b) normalized purity $\Gamma$, for a qubit state probed with the tetrahedron
  measurement.
  The red curves labeled `a' are for the primitive prior; 
  the blue curves labeled `b' are for the Jeffreys prior.
  These curves delineate the boundaries of the SCIs for different credibility
  values; the cusps are located at the maximum-likelihood estimates
  $\widehat{\Phi}_{\textsc{ml}}$ and $\widehat{\Gamma}_{\textsc{ml}}$, respectively.
  For illustration, the plausible intervals for the primitive prior, which are
  the SCIs with respective credibility $0.932$ and $0.697$, are
  indicated by the black bars.
  The true values of $\Phi=0.9747$ and $\Gamma=0.81$, marked by the
  down-pointing arrows ($\downarrow$), happen to be inside these
  SCIs.
  Although, only ${N=36}$ qubits are measured in the simulated experiment, the
  SCIs are almost the same for the two priors.}
\end{figure}

We construct the SCIs for two properties: the fidelity with respect to some
target state, and the normalized purity.
Both have values between $0$ and $1$, so that the concepts and tools of
the preceding sections are immediately applicable.

The fidelity
\begin{equation}\label{eq:7-0a}
  \phi=\tr{\bfsym{|}\sqrt{\rho}\,\sqrt{\rho_{\mathrm{tar}}}\bfsym{|}}
\end{equation}
is a measure of overlap between the actual state $\rho$ and the target
state $\rho_{\mathrm{tar}}$.
For these two qubit states, we express the fidelity in terms of the Bloch
vectors $\vec{r}$ and ${\vec{t}=\tr{\bfsym{\sigma}\rho_{\mathrm{tar}}}}$,
\begin{equation}\label{eq:7-0b}
  \phi={\left[\frac{1}{2}(1+\vec{r}\cdot\vec{t})
+\frac{1}{2}\sqrt{1-r^2}\sqrt{1-t^2}\right]}^{\frac{1}{2}}
\geq\sqrt{\frac{1-t}{2}}\,,
\end{equation}
where ${r=\bfsym{|}\vec{r}\bfsym{|}}$ and ${t=\bfsym{|}\vec{t}\bfsym{|}}$,
and the lower bound is reached for ${\vec{r}=-\vec{t}/t}$.
When the target state is pure
($\rho_{\mathrm{tar}}=|\mathrm{tar}\rangle\langle\mathrm{tar}|$,
${t=1}$), (\ref{eq:7-0b}) simplifies to
$\phi
={\left[\frac{1}{2}(1+\vec{r}\cdot\vec{t})\right]}^{\frac{1}{2}}
=\langle\mathrm{tar}|\rho|\mathrm{tar}\rangle^{\frac{1}{2}}$.
In particular, for
$|\mathrm{tar}\rangle=|0\rangle$, the $+1$ eigenstate of $\sigma_z$,
we have ${\vec{t}=\vec{e}_z}$, and the fidelity is a function of only the
$z$-component of $\vec{r}$, namely
$\phi={\left[\frac{1}{2}(1+z)\right]}^{\frac{1}{2}}$.

The purity $\tr{\rho^2}$ is a measure of the mixedness of a state, with
values between $\frac{1}{2}$ (for the completely mixed state) and $1$ (for
pure states).
We define the normalized purity by ${\gamma=2\,\tr{\rho^2}-1}$, so that
${\gamma=r^2}$ is simply the squared length of the Bloch vector.
Expressed in terms of the tetrahedron probabilities in Eq.~\eqref{7-5}, we have
\begin{equation}\label{eq:7-0d}
  \gamma(p)=12\sum_{k=1}^4p_k^2-3
\quad\mbox{and}\quad
\phi(p)=\sqrt{1-2p_1}
\end{equation}
for the normalized purity and the fidelity with
$\rho_{\mathrm{tar}}={\frac{1}{2}(1+\sigma_z)}$, respectively.

In a simulated experiment, the state used to generate the data is
${\rho=\frac{1}{2}(\identity+0.9\,\sigma_{z})}$.
This state has fidelity ${\Phi=\sqrt{0.95}=0.9747}$ (for target state
$|0\rangle$) and normalized purity ${\Gamma=0.81}$ --- the ``true'' values
for the two properties to be estimated from the data.
A particular simulation measured $36$ copies of this state using the
tetrahedron measurement, and gave data $D=(n_1,n_2,n_3,n_4)=(2,10,11,13)$,
where $n_k$ is the number of clicks registered by the detector for
outcome $\Pi_k$.

In this low-dimensional single-qubit case, the induced priors $W_0(\Phi)$ and
$W_0(\Gamma)$, both for the primitive prior \eqref{4-6} and the Jeffreys prior
\eqref{4-7}, are obtained by an analytical evaluation of the analogs of the
integral in Eq.~\eqref{3-4}.
While a MC integration is needed for the analogs of the integral in
Eq.~\eqref{3-5}, one can do without the full machinery of Sec.~\ref{sec:MCint}.
The top plots in Fig.~\ref{fig:Size&Credibility} report the $F$-likelihoods
$L(D|\Phi)$ and $L(D|\Gamma)$ thus obtained for the Jeffreys prior and the
primitive prior, respectively.

The bottom plots show the size $s_{\lambda}^{\ }$ and the credibility
$c_{\lambda}^{\ }$ for the resulting BLIs, computed from these $F$-likelihoods
together with the respective induced priors.
The dots mark values obtained by numerical integration that employs
the Hamiltonian Monte Carlo algorithm for sampling the quantum state
space \cite{Seah+4:15} in accordance with the prior and posterior
distributions.
Consistency with the relation in Eq.~\eqref{5-7} between $s_\lambda$ and
$c_\lambda$ is demonstrated by the green curves through the credibility
points, which is obtained by integrating over the cyan curve fitted to the
size points.

The SCIs resulting from these $s_\lambda$ and $c_\lambda$ are reported in
Fig.~\figref{qubitSCIs} for
fidelity $\Phi$ and normalized purity $\Gamma$, both for the primitive
prior (red lines `a') and for the Jeffreys prior (blue lines `b').
The SCI with a specific credibility is the horizontal
interval between the two branches of the curves;
see the plausible intervals marked on the plots.
An immediate observation is that the choice of prior has little effect
on the SCIs, although the total number of measured copies is not large.
In other words, already for the small number of ${N=36}$ qubits measured, our
conclusions are dominated by the data, not by the choice of prior.

\begin{figure}[!t]\centering
  \includegraphics{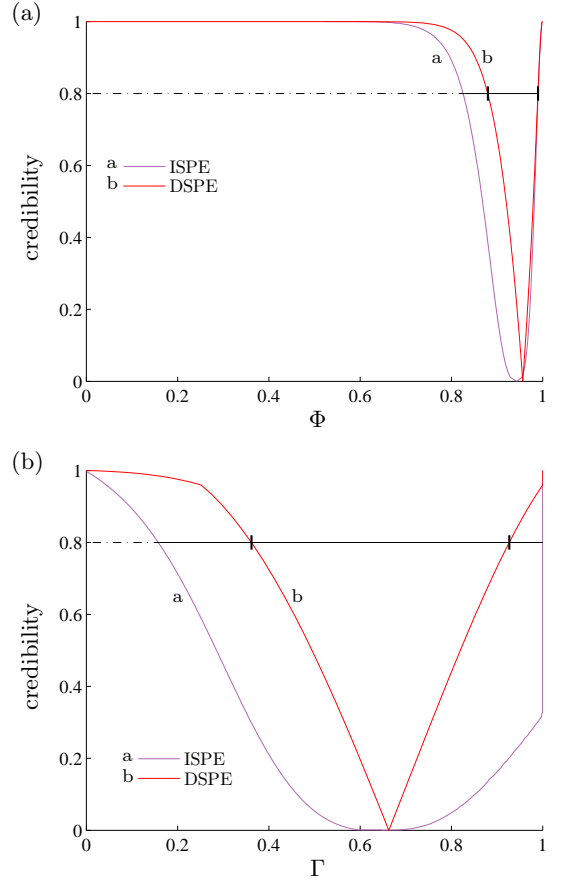}
  \caption{\label{fig:DSPEvsISPE}%
  Direct and indirect state-property estimation:
  Error intervals for (a) fidelity $\Phi$ and (b) normalized purity $\Gamma$
  by ISPE (purple curves `a') and DSPE (red curves `b'),
  for the same simulated data as in Figs.~\figref{Size&Credibility} and
  \figref{qubitSCIs}.
  The horizontal lines indicate the intervals
  for credibility $0.8$ --- credibility of the interval in the case of DSPE
  but credibility for the reconstruction-space region in the case of ISPE.
  Consistent with what the sketch in Fig.~\figref{regions} suggests,
  the intervals obtained from ISPE are larger than the actual SCIs that result
  from proper DSPE.
  In plot (a), one can also clearly see that the maximum-likelihood
  fidelity $\widehat{\Phi}_{\textsc{ml}}$ is not the fidelity of the
  maximum-likelihood state $\rhoML$:
  The cusps of the red and purple curves are at different $\Phi$ values.}
\end{figure}

\subsection{Direct and indirect estimation of state properties}\label{sec:1qubitb}
As mentioned in the Introduction and also in Sec.~\ref{sec:PE-OEI},
the best guess for the properties of interest may not, and often does
not, come from the best guess for the quantum state.
For an illustration of this matter, we compare here the two approaches for our
qubit example.
The error intervals are either constructed by directly estimating the value of
the property from the data, as we have done in the previous section, or by
first constructing the error regions (SCRs specifically; see
Ref.~\cite{Shang+4:13}) for the quantum state, and the error interval for the
desired property is given by the range of property values for the states
contained in the error region of states; see Fig.~\figref{regions}.
We refer to the two respective approaches as direct and indirect state-property
estimation, with the abbreviations of DSPE and ISPE.
Of course, DSPE is simply SPE proper.

Figure~\figref{DSPEvsISPE} shows the error intervals for fidelity
$\Phi$ and normalized purity $\Gamma$ for the single-qubit data of
Figs.~\figref{Size&Credibility} and \figref{qubitSCIs}.
The purple curves labeled `a' are obtained via ISPE and the red curves 
labeled `b' via DSPE. 
Here, the primitive prior of Eq.~\eqref{4-6} is used as $w_0(p)$ on the
probability space, together with the induced prior densities $W_0(\Phi)$ and
$W_0(\Gamma)$ for the fidelity and the normalized purity.
Clearly, the error intervals obtained by these two approaches are
quite different in this situation and, in particular,
DSPE reports smaller intervals than ISPE does.
More importantly, the intervals obtained via ISPE and DSPE are also rather
different in meaning:
The credibility value used for constructing the interval from DSPE (the SCI)
is the posterior content of that interval for the property itself; the
credibility value used in ISPE, however, is the posterior content for the
\emph{state} error region, and often has no simple relation to the probability
of containing the true property value.
This is the situation depicted in Fig.~\figref{regions}, where the range
of $F$ values across the SCR is larger than the range of the SCI.

\section{Example: Two qubits}\label{sec:2qubit}
\subsection{CHSH quantity, TAT scheme, and simulated experiment}
In our second example we consider qubit pairs and, as in Sec.~4.3 in
Ref.~\cite{Seah+4:15}, the property of interest is the
Clauser-Horne-Shimony-Holt (CHSH) quantity \cite{Clauser+3:69,Clauser+1:74},%
\begin{equation}\label{eq:8-1}
  \chsh=\tr{(A_1\otimes B_1 + A_2\otimes B_1 + A_1\otimes B_2
             - A_2\otimes B_2)\rho}\,,
\end{equation}
where $A_j=\vec{a}_j\cdot\bfsym{\sigma}$ and
$B_{j'}=\vec{b}_{j'}\cdot\bfsym{\sigma}$ with unit vectors $\vec{a}_1$,
$\vec{a}_2$, $\vec{b}_1$, and $\vec{b}_2$ are components of the Pauli vector
operators for the two qubits.
We recall that $\bfsym{|}\chsh\bfsym{|}$ cannot exceed $\sqrt{8}$, and the
two-qubit state is surely entangled if ${\bfsym{|}\chsh\bfsym{|}>2}$.
Therefore, one usually wishes to distinguish reliably between
${\bfsym{|}\CHSH\bfsym{|}<2}$ and ${\bfsym{|}\CHSH\bfsym{|}>2}$.

A standard choice for the single-qubit observables is
\begin{equation}\label{eq:8-2}
  A_1=\sigma_x\,,\quad A_2=\sigma_z\,,\quad
\left.\begin{array}{c}B_1\\B_2\end{array}\right\}=
-\frac{1}{\sqrt{2}}(\sigma_x\pm\sigma_z)\,,
\end{equation}
for which
\begin{equation}\label{eq:8-3}
  \chsh=-\sqrt{2}\expect{\sigma_x\otimes\sigma_x+\sigma_z\otimes\sigma_z}\,.
\end{equation}
The limiting values ${\chsh=\pm\sqrt{8}}$ are reached for
two of the ``Bell states'',
\textsl{viz.}\ the maximally entangled states
$\rho=\frac{1}{4}(\identity\mp\sigma_x\otimes\sigma_x)%
(\identity\mp\sigma_z\otimes\sigma_z)$, the common eigenstates of
$\sigma_x\otimes\sigma_x$ and $\sigma_z\otimes\sigma_z$ with same eigenvalue
$-1$ or $+1$.

One does not need  full tomography for the experimental determination of this
$\CHSH$;
a measurement that explores the $xz$ planes of the two Bloch balls
provides the necessary data.
We use the trine-antitrine (TAT) scheme (see Ref.~\cite{Tabia+1:11} and Sec.~6
in Ref.~\cite{Shang+4:13}) for this purpose.
Qubit~1 is measured by the three-outcome POM with
outcome operators
\begin{equation}\label{eq:8-4}
  \Pi_1^{(1)}=\frac{1}{3}\left(\mathds{1}+\sigma_z\right),\quad
\left.\begin{array}{c}\Pi_2^{(1)}\\[0.5ex]\Pi_3^{(1)}\end{array}\right\}=
\frac{1}{3}{\left(\mathds{1}\pm
    \frac{\sqrt{3}}{2}\sigma_x-\frac{1}{2}\sigma_z\right)},
\end{equation}
and the $\Pi_{j'}^{(2)}$s for qubit~2 have the signs of $\sigma_x$ and
$\sigma_z$ reversed.
The nine probability operators of the product POM are
\begin{equation}\label{eq:8-5}
  \Pi_k^{\ }=\Pi_{j}^{(1)}\otimes\Pi_{j'}^{(2)}\quad\mbox{with}\quad
  k=3(j-1)+j'\equiv[jj']\,,
\end{equation}
that is $1=[11]$, $2=[12]$, \dots, $5=[22]$, \dots, $8=[32]$, $9=[33]$,
and we have
\begin{equation}\label{eq:8-6}
   \chsh(p)=\sqrt{8}\Bigl[3(p_1+p_5+p_9)-1\Bigr]
\end{equation}
for the CHSH quantity in Eq.~\eqref{8-3}.

With the data provided by the TAT measurement, we can evaluate $\chsh$ for any
choice of the unit vectors $\vec{a}_1$, $\vec{a}_2$, $\vec{b}_1$, $\vec{b}_2$
in the $xz$ plane.
If we choose the vectors such that $\chsh$ is largest for the given $\rho$, then
\begin{eqnarray}\label{eq:8-7}
  \chshopt&=&2\Bigl[\expect{\sigma_x\otimes\sigma_x}^2
                       +\expect{\sigma_x\otimes\sigma_z}^2
                       +\expect{\sigma_z\otimes\sigma_x}^2\nonumber\\
  &&\phantom{2\Bigl[}+\expect{\sigma_z\otimes\sigma_z}^2\Bigr]^{\frac{1}{2}}
\end{eqnarray}
for the optimized CHSH quantity.
In terms of the TAT probabilities, it is given by
\begin{eqnarray}\label{eq:8-8}
  {\left(\frac{\chshopt(p)}{4}\right)}^2&=&
 1+9\sum_{k=1}^9p_k^2\\&&\mbox{}
 -3\Bigl[(p_1+p_2+p_3)^2+(p_4+p_5+p_6)^2\nonumber\\&&\phantom{-3\Bigl[}
 +(p_7+p_8+p_9)^2+(p_1+p_4+p_7)^2\nonumber\\
 &&\phantom{-3\Bigl[}+(p_2+p_5+p_8)^2+(p_3+p_6+p_9)^2\Bigr].\nonumber
\end{eqnarray}
Whereas the fixed-vectors CHSH quantity in Eq.~\eqref{8-6} is a linear
function of the TAT probabilities, the optimal-vectors quantity is not.

The inequality $\bfsym{|}\chsh\bfsym{|}\leq \chshopt$ holds for any
two-qubit state $\rho$, of course.
Extreme examples are the Bell states
$\rho=\frac{1}{4}(\identity\pm\sigma_x\otimes\sigma_x)%
(\identity\mp\sigma_z\otimes\sigma_z)$, the common eigenstates of
$\sigma_x\otimes\sigma_x$ and $\sigma_z\otimes\sigma_z$ with opposite
eigenvalues, for which ${\chsh=0}$ and ${\chshopt=\sqrt{8}}$.
The same values are also found for other states, among them all four common
eigenstates of $\sigma_x\otimes\sigma_z$ and $\sigma_z\otimes\sigma_x$.

The simulated experiment uses the true state
\begin{equation}\label{eq:8-9}
  \rho^{\ }_{\mathrm{true}}=\frac{1}{4}(\identity-x\sigma_x\otimes\sigma_x
                                                 -y\sigma_y\otimes\sigma_y
                                                 -z\sigma_z\otimes\sigma_z)
\end{equation}
with $(x,y,z)=\frac{1}{20}(18,-15,-14)$, for which the TAT probabilities are
\begin{equation}\label{eq:8-10}
  {\left(\begin{array}{ccc}
  p_1 & p_2 & p_3 \\
  p_4 & p_5 & p_6 \\
  p_7 & p_8 & p_9
  \end{array}\right)}=\frac{1}{60}
  {\left(\begin{array}{ccc}
    2 & 9 & 9 \\ 9 & 10 & 1 \\ 9 & 1 & 10
  \end{array}\right)}
\end{equation}
and the true values of $\CHSH$ and $\CHSHopt$ are
\begin{eqnarray}\label{eq:8-11}
  &&\CHSH=\sqrt{2}(x+z)=\frac{1}{5}\sqrt{2}=0.2828\,,\nonumber\\
  &&\CHSHopt=2\sqrt{x^2+z^2}=\sqrt{\frac{26}{5}}=2.2804\,.
\end{eqnarray}
When simulating the detection of ${N=180}$ copies, we obtained the relative
frequencies
\begin{eqnarray}\label{eq:8-12}
  \frac{1}{180}
  {\left(\begin{array}{ccc}
    9 & 28 & 30 \\ 28 & 27 & 3 \\ 29 & 1 & 25
  \end{array}\right)}
 &=&  {\left(\begin{array}{ccc}
  p_1 & p_2 & p_3 \\
  p_4 & p_5 & p_6 \\
  p_7 & p_8 & p_9
  \end{array}\right)}\nonumber\\&&\mbox{}+\frac{1}{180}
  {\left(\begin{array}{ccc}
    3 & 1 & 3 \\ 1 & -3 & 0 \\ 2 & -2 & -5
  \end{array}\right)}.
\end{eqnarray}
If we estimate the probabilities by the relative frequencies
and use these estimates in Eqs.~\eqref{8-6} and \eqref{8-8}, the resulting
estimates for $\CHSH$ and $\CHSHopt$ are ${\sqrt{2}/30=0.0471}$ and
${16\sqrt{39}/45=2.2204}$, respectively.

This so-called ``linear inversion'' is popular, and one can supplement the
estimates with error bars that refer to confidence intervals
\cite{Schwemmer+6:15}, but the approach has well-known problems
\cite{Shang+2:14}.
Instead, we report SCIs for $\CHSH$ and $\CHSHopt$, and for those we need the
$\CHSH$-likelihoods $L(D|\CHSH)$ and $L(D|\CHSHopt)$.
We describe in the following Sec.~\ref{sec:2qb-MCint} how the iteration
algorithm of Sec.~\ref{sec:MCint} is implemented, and present $L(D|\CHSH)$
and $L(D|\CHSHopt)$ thus found in Sec.~\ref{sec:2qb-SCIs} together with the
resulting SCIs.

\subsection{Iterated MC integrations}\label{sec:2qb-MCint}
Rather than ${F=\frac{1}{2}{\left(\CHSH/\sqrt{8}+1\right)}}$ or
${F=\CHSHopt/\sqrt{8}}$, which have values in the range
${0\leq F\leq1}$, we shall use $\CHSH$ and
$\CHSHopt$ themselves as the properties to be estimated,
with the necessary changes in the expressions
in Secs.~\ref{sec:SCPR}--\ref{sec:MCint}.
For the MC integration of $P_0(\CHSH)$, say, we sample the probability space
with the Hamiltonian MC algorithm described in Sec.~4.3 in
\cite{Seah+4:15}.

In this context, we note the following implementation issue:
The sample probabilities carry a weight proportional to the range of
permissible values for
${\expect{(\sigma_x\otimes\sigma_x)(\sigma_z\otimes\sigma_z)}
=-\expect{\sigma_y\otimes\sigma_y}}$, i.e., parameter $q$ in \eqref{B-4}.
It is expedient to generate an unweighted sample by resampling
(``bootstrapping'') the weighted sample.
The unweighted sample is then used for the MC integration.

\begin{figure}[!t]\centering
  \includegraphics{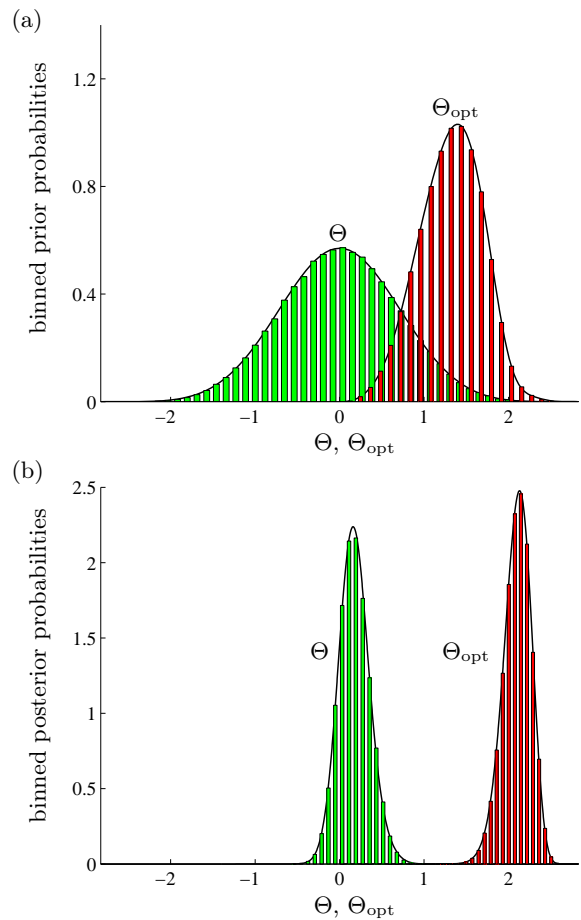}
    \caption{\label{fig:CHSH-histo}%
    (a) Histogram of CHSH values in a random sample of 500\,000 states in
    accordance with the primitive prior of Eq.~\eqref{4-6}.
    For $\CHSH$ of Eq.~\eqref{8-6} we have the full range of
    ${-\sqrt{8}\leq\CHSH\leq\sqrt{8}}$, whereas $\CHSHopt$ of Eq.~\eqref{8-8}
    is positive by construction. ---
    (b) Corresponding histogram for a random sample drawn from the
    posterior distribution for the simulated data in Eq.~\eqref{8-12}.
    --- In plot (a), the black-line envelopes show the few-parameter
    approximations of Eq.~\eqref{8-14} with Eq.~\eqref{8-15} for $\CHSH$ and
    Eqs.~\eqref{B-17}--\eqref{B-19} for $\CHSHopt$.
    In plot (b), the envelopes are the derivatives of the fits to
    $\prefD(\CHSH)$ and $\prefD(\CHSHopt)$.}
\end{figure}

The histograms in Fig.~\figref{CHSH-histo}(a) show the distribution
of $\CHSH$ and $\CHSHopt$ values in such a sample, drawn from the
probability space in accordance with the primitive prior of \eqref{4-6}.
These prior distributions contain few values with ${\CHSHopt>2}$ and
much fewer with ${\bfsym{|}\CHSH\bfsym{|}>2}$.
In Fig.~\figref{CHSH-histo}(b), we have the histograms for a corresponding
sample drawn from
the posterior distribution to the simulated data of Eq.~\eqref{8-12}.
In the posterior distributions,
values exceeding $2$ are prominent for $\CHSHopt$, but
virtually non-existent for~$\CHSH$.

We determine the $\CHSH$-likelihoods $L(D|\CHSH)$ and $L(D|\CHSHopt)$ by the
method described in Sec.~\ref{sec:MCint}.
The next five paragraphs deal with the details of carrying out a
few rounds of the iteration.

\begin{figure}[!t]\centering
  \includegraphics{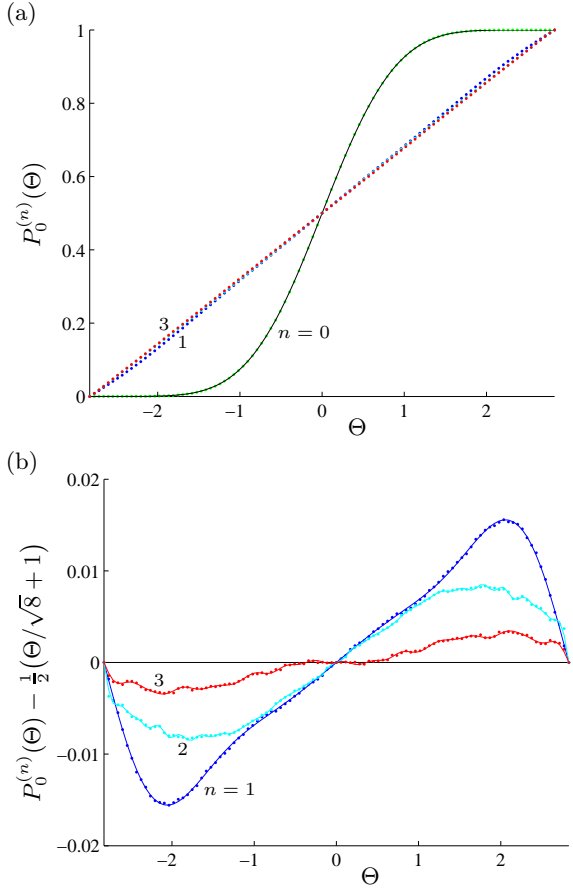}
    \caption{\label{fig:P0-Iteration}%
    Consecutive functions $P^{(n)}_0(\CHSH)$ for $n=0,1,2,3$ as obtained by MC
    integration.
    The green dots (${n=0}$) represent values for $P_0(\CHSH)$, computed
    with the primitive prior \eqref{4-6}.
    The flat regions near the end points at ${\CHSH=\pm\sqrt{8}}$ are a
    consequence of the $\frac{11}{2}$ power in Eq.~\eqref{8-13}.
    The black curve through the green dots is the graph of the
    four-parameter approximation $P^{(0)}_0(\CHSH)$ of Eq.~\eqref{8-14}.
    The blue, cyan, and red dots are the MC values for ${n=1}$, $2$, and $3$,
    respectively, all close to the straight line
    ${\CHSH\mapsto\frac{1}{2}{\left(\CHSH/\sqrt{8}+1\right)}}$.
    The cyan dots are difficult to see between the blue and red
    dots in plot (a). They are well visible in plot (b), where the
    straight-line values are subtracted.
    The curves through the dots in plot (b) show the few-term Fourier
    approximations analogous to Eq.~\eqref{6-7}.
}
\end{figure}

The green dots in Fig.~\figref{P0-Iteration}(a) show the $P_0(\CHSH)$ values
obtained with the sample of 500\,000 sets of probabilities that generated
the histograms in Fig.~\figref{CHSH-histo}(a).
We note that the MC integration is not precise enough to distinguish
${P_0(\CHSH)\gtrsim0}$ from ${P_0(\CHSH)=0}$ for ${\CHSH<-2}$ or
${P_0(\CHSH)\lesssim1}$ from ${P_0(\CHSH)=1}$ for ${\CHSH>2}$ and, therefore,
we cannot infer a reliable approximation for
${W_0(\CHSH)=\frac{\D}{\D\CHSH}P_0(\CHSH)}$ for these $\CHSH$ values;
the sample contains only 144 entries with
${\bfsym{|}\CHSH\bfsym{|}>2}$ and no entries with
${\bfsym{|}\CHSH\bfsym{|}>2.49}$.
The iteration algorithm solves this problem.

As discussed in Appendix~\ref{sec:appA}, we have
\begin{equation}\label{eq:8-13}
  \frac{\D}{\D\CHSH}P_0(\CHSH)=W_0(\CHSH)\propto
        {\left(\sqrt{8}-\bfsym{|}\CHSH\bfsym{|}\right)}^\frac{11}{2}
\quad\mbox{for}\quad
\bfsym{|}\CHSH\bfsym{|}\lesssim\sqrt{8}
\end{equation}
near the boundaries of the $\CHSH$ range in Fig.~\figref{P0-Iteration}(a).
In conjunction with the symmetry property ${W_0(\CHSH)=W_0(-\CHSH)}$ or
${P_0(\CHSH)+P_0(-\CHSH)=1}$, this invites the four-parameter approximation
\begin{eqnarray}\label{eq:8-14}
  P_0(\CHSH)\simeq P^{(0)}_0(\CHSH)&=& w_1B_{\alpha_1^{\ }}(\CHSH)
                                  +w_2B_{\alpha_2^{\ }}(\CHSH)\nonumber\\
                         &&\mbox{}+w_3B_{\alpha_3^{\ }}(\CHSH)\,,
\end{eqnarray}
where
\begin{equation}\label{eq:8-15}
  B_{\alpha}(\CHSH)=\Biggr(\frac{1}{32}\Biggr)^{\alpha+\frac{1}{2}}
              \frac{(2\alpha+1)!}{(\alpha!)^2}
                \int_{-\sqrt{8}}^{\CHSH}\D x\,{\left(8-x^2\right)}^{\alpha}
\end{equation}
is a normalized incomplete beta function integral with
${B_{\alpha}(-\sqrt{8})=0}$ and ${B_{\alpha}(\sqrt{8})=1}$;
$\alpha_2$ and $\alpha_3$ are fitting parameters larger than
${\alpha_1=\frac{11}{2}}$; and $w_1,w_2,w_3$ are weights with unit sum.
A fit with a root mean squared error of $2.7\times10^{-4}$ is achieved by
$\alpha_2=\alpha_1+1.6700$, $\alpha_3=\alpha_1+5.4886$,
and $(w_1,w_2,w_3)=(0.4691,0.2190,0.3119)$.
The graph of $P^{(0)}_0(\CHSH)$ is the black curve through the green dots in
Fig.~\figref{P0-Iteration}(a); the corresponding four-parameter
approximation for $W_0(\CHSH)$ is shown as the black envelope for the green
$\CHSH$ histogram in Fig.~\ref{fig:CHSH-histo}(a).

The subsequent approximations $P_0^{(1)}(\CHSH)$, $P_0^{(2)}(\CHSH)$, and
$P_0^{(3)}(\CHSH)$, are shown as the blue, cyan, and red dots in
Fig.~\figref{P0-Iteration}(a) and, after subtracting
$\frac{1}{2}{\left(\CHSH/\sqrt{8}+1\right)}$, also in Fig.~\figref{P0-Iteration}(b).
We use the truncated Fourier series of Eq.~\eqref{6-7} with
${F=\frac{1}{2}{\left(\CHSH/\sqrt{8}+1\right)}}$ for fitting a smooth curve to
the noisy MC values for $P_0^{(1)}(\CHSH)$, $P_0^{(2)}(\CHSH)$, and
$P_0^{(3)}(\CHSH)$.
As a consequence of ${P_0(\CHSH)+P_0(-\CHSH)=1}$, all Fourier amplitudes $a_k$
with odd $k$ vanish.

\begin{figure}\centering
  \includegraphics{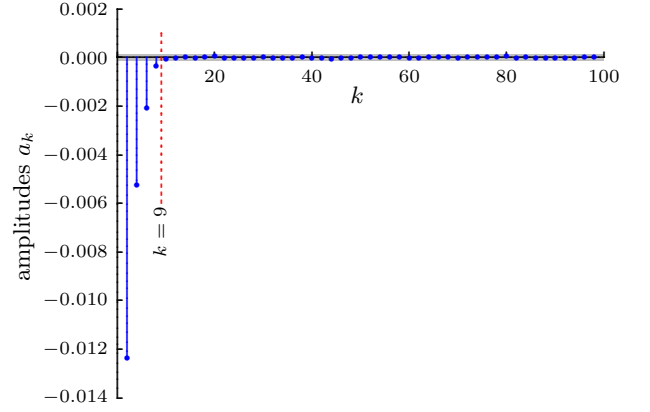}
    \caption{\label{fig:Fourier}%
    Fourier coefficients of Eq.~\eqref{6-7} for $P_0^{(1)}(\CHSH)$ ($\equiv$ blue
    dots in Fig.~\figref{P0-Iteration}).
    All amplitudes with odd index vanish, ${a_1=a_3=a_5=\cdots=0}$, and are
    not included in the figure.
    The ``low-pass filter'' set at ${k=9}$ keeps only the four amplitudes
    $a_2$, $a_4$, $a_6$, and $a_8$ in order to remove the high-frequency noise
    in $P_0^{(1)}(\CHSH)$.
    Each of the discarded amplitudes is less than $1\%$ in magnitude of the
    largest amplitude $a_2$; the gray strip about the horizontal axis
    indicates this $1\%$ band.}
\end{figure}

For an illustration of the method, we report in Fig.~\figref{Fourier} the
amplitudes $a_k$ of a full Fourier interpolation between the blue dots
(${n=1}$) in Fig.~\figref{P0-Iteration}(b).
Upon discarding all components with ${k>8}$ and thus retaining only
four nonzero amplitudes, the resulting truncated Fourier series gives
the smooth blue curve through the blue dots.
Its derivative contributes a factor $W_0^{(1)}(F)$ to the reference prior
density $W_{\mathrm{r},0}(F)$, in accordance with step S5 of the iteration
algorithm in Sec.~\ref{sec:MCint}.
In the next round we treat $P_0^{(2)}(\CHSH)$ in the same way, followed by
$P_0^{(3)}(\CHSH)$ in the third round.

\subsection{Likelihood and optimal error intervals}\label{sec:2qb-SCIs}
\begin{figure}\centering
  \includegraphics{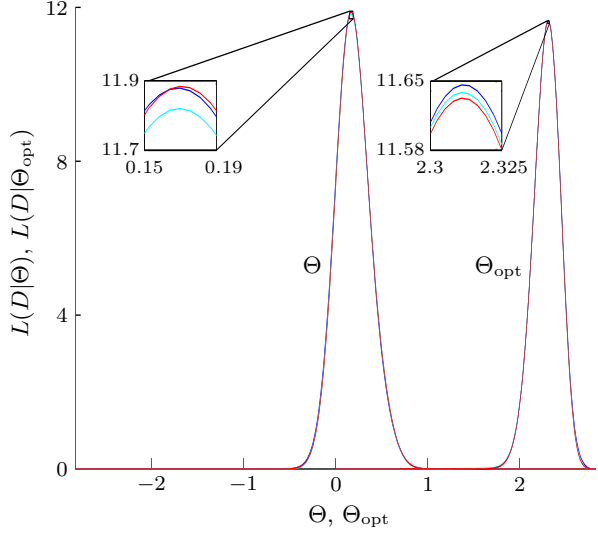}
    \caption{\label{fig:S-Likelihood}%
    Likelihood function for $\CHSH$ and $\CHSHopt$.
    The plot of $L(D|\CHSH)$ shows the $\CHSH$-likelihood obtained for the three
    subsequent iterations in Fig.~\figref{P0-Iteration}(b), with a blow-up
    of the region near the maximum.
    The colors blue, cyan, and red correspond to those in
    Fig.~\figref{P0-Iteration}. ---
    The plot of $L(D|\CHSHopt)$ is analogous.}
\end{figure}

After each iteration round, we use the current reference prior and the
likelihood $L(D|p)$ for a MC integration of the posterior density and so
obtain the corresponding $P_D^{(n)}(\CHSH)$ as well as its analytical
parameterization analogous to that of $P_0^{(n)}(\CHSH)$; the black
envelopes to the histograms in Fig.~\ref{fig:CHSH-histo}(b) show the
final approximations for the derivatives of $\prefD(\CHSH)$ and
$\prefD(\CHSHopt)$ thus obtained.
The ratio of their derivatives is the $n$th approximation to the
$\CHSH$-likelihood $L(D|\CHSH)$; and likewise for $L(D|\CHSHopt)$, see
\ref{sec:appB}.
Figure~\figref{S-Likelihood} shows the sequence of approximations.

\begin{figure}\centering
  \includegraphics{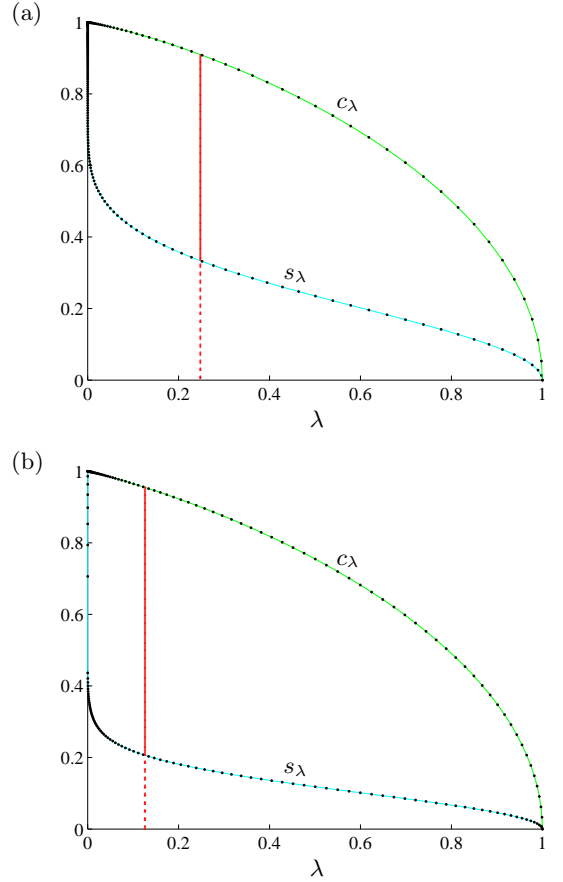}
    \caption{\label{fig:BLI-SizeCred}%
    Size and credibility of bounded-likelihood intervals for the CHSH
    quantities, computed from the likelihood functions in
    Fig.~\figref{S-Likelihood}.
    (a) Fixed measurement of Eq.~\eqref{8-6} with the flat prior density
    ${W_0(\CHSH)=1/\sqrt{32}}$;
    (b) optimized measurement of Eq.~\eqref{8-8} with the flat prior density
    ${W_0(\CHSHopt)=1/\sqrt{8}}$.
    The red vertical lines mark the critical $\lambda$ values at
    $\lamcr=0.2488$ and $\lamcr=0.1267$.}
\end{figure}

We note that the approximations for the $\CHSH$-likelihood hardly change from
one iteration to the next, so that we can stop after just a few rounds and
proceed to the calculation of the size $s_\lambda$ and the credibility
$c_\lambda$ of the BLIs.
These are shown in Fig.~\figref{BLI-SizeCred} for the flat priors in $\CHSH$
and $\CHSHopt$, respectively.

\begin{figure}\centering
  \includegraphics{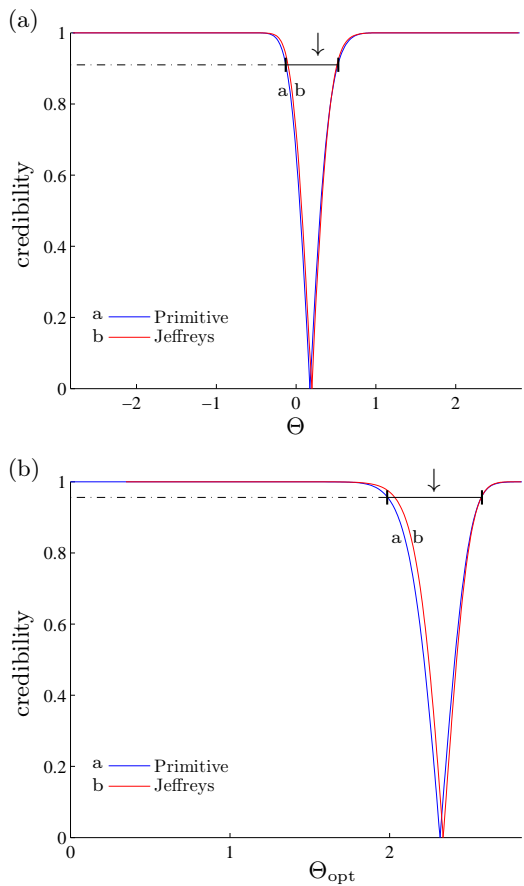}
    \caption{\label{fig:S-OEIs}%
    Optimal error intervals for (a) $\CHSH$ and (b) $\CHSHopt$.
    The blue and red curves (labeled `a' and `b', respectively)
    delineate the boundaries of the SCIs in the same
    manner as in Figs.~\figref{qubitSCIs} and \figref{DSPEvsISPE}.
    The true values of ${\CHSH=0.0471}$ and ${\CHSHopt=2.2204}$,
    marked by the down-pointing arrows ($\downarrow$),
    are inside the indicated plausible intervals for the primitive reference
    prior with credibility $0.910$ and $0.956$, respectively.
    --- The primitive prior of Eq.~\eqref{4-6} and the Jeffreys prior of
    Eq.~\eqref{4-7} solely serve as the reference priors on the probability
    space for the computation of the $\CHSH$-likelihoods (shown in
    Fig.~\ref{fig:S-Likelihood} for the primitive prior),
    whereas flat priors for
    $\CHSH$ and $\CHSHopt$ are used for establishing the boundaries of the
    SCIs from these $\CHSH$-likelihoods.}
\end{figure}

The plots in Figs.~\ref{fig:CHSH-histo}--\ref{fig:BLI-SizeCred} refer to
the primitive prior of Eq.~\eqref{4-6} as the reference prior on the probability
space.
The analogous plots for the Jeffreys prior of Eq.~\eqref{4-7} are quite similar.
As a consequence of this similarity, there is not much of a difference in the
SCIs obtained for the two reference priors, although the number of
measured copies (${N=180}$) is not large; see Fig.~\figref{S-OEIs}.
The advantage of $\CHSHopt$ over $\CHSH$ is obvious:
Whereas virtually all $\CHSH$-SCIs with non-unit credibility are inside the
range ${-2<\CHSH<2}$, 
the $\CHSHopt$-SCIs are entirely in the range ${\CHSHopt>2}$
for credibility up to 95\% and 98\% for the primitive reference prior
and the Jeffreys reference prior, respectively.

\section{Summary and outlook}\label{sec:Sum}
In full analogy to the likelihood $L(D|p)$ of the data $D$ for the specified
probability parameters $p$ of the quantum state, which is the basic ingredient
exploited by all strategies for quantum state estimation, the $F$-likelihood
$L(D|F)$ plays this role when one estimates the value $F$ of a function $f(p)$
--- the value of a property of the quantum state.
Although the definition of $L(D|F)$ in terms of $L(D|p)$ relies on Bayesian
methodology and, in particular, needs a pre-selected reference prior on the
probability space, the prior density for $F$ can be chosen freely and the
$F$-likelihood is independent of this choice.

As soon as the $F$-likelihood is at hand, we have a maximum-likelihood
estimator for $F$, embedded in a family of smallest credible intervals that
report the accuracy of the estimate in a meaningful way.
This makes optimal use of the data.
The dependence of the smallest credible regions on the prior density for $F$
is irrelevant when enough data are available.
In the examples studied, ``enough data'' are obtained by measuring a few
tens of copies per outcome.

Not only is there no need for estimating the quantum state first and finding
its smallest credible regions, this is not even useful:
The $F$ value of the best-guess state is not the best guess for $F$,
and the smallest credible region for the state does not carry the meaning of
the smallest credible interval for $F$.

The reliable computation of the marginal $F$-likelihood $L(D|F)$ from
the primary state-conditioned likelihood $L(D|p)$ is indeed possible.
It requires the evaluation of
high-dimensional integrals with Monte Carlo techniques.
It can easily happen that the pre-selected prior on the probability space
gives very little weight to sizeable ranges of $F$ values, and then the
$F$-likelihood is ambiguous there.
We overcome this problem by an iterative algorithm that replaces the
inadequate prior by suitable ones, and so yields a $F$-likelihood that is
reliable for all values of $F$.
The two-qubit example, in which we estimate CHSH quantities, illustrates these
matters.

From a general point of view, one could regard values $F$ of functions $f(p)$
of the quantum state as parameters of the state.
The term \emph{quantum parameter estimation} is, however, traditionally used for
the estimation of parameters of the experimental apparatus, such as the
emission rate of the source, efficiencies of detectors, or the phase of an
interferometer loop.
A forthcoming paper \cite{Dai+4:16} will deal with optimal error regions for
quantum parameter estimation in this traditional sense --- smallest credible
regions, that is.
In this context, it is necessary to account, in the proper way, for the
quantum systems that are emitted by the source but escape detection.

There are also situations, in which the quantum state and parameters of the
apparatus are estimated from the same data, often referred to as
\emph{self-calibrating experiments} \cite{Mogilevtsev+2:09,Mogilevtsev:10}.
Various aspects of the combined optimal error regions for the parameters of
both kinds are discussed in~\cite{Sim:15} and are the subject matter of
ongoing research.

\begin{acknowledgments}
We thank David Nott and Michael Evans for stimulating discussions.
This work is funded by the Singapore Ministry of Education (partly through the
Academic Research Fund Tier 3 MOE2012-T3-1-009) and the National Research
Foundation of Singapore.
H.~K.~N is also funded by a Yale-NUS College start-up grant.  
\end{acknowledgments}

\appendix
\renewcommand{\thesubsection}{\Alph{section}.\arabic{subsection}}

\section{Confidence \textit{vs} credibility}\label{sec:appCC}
It is common practice to state the result of a measurement of a physical
quantity in terms of a \emph{confidence interval}.
Usually, two standard deviations on either side of the average value define a
95\% confidence interval for the observed data.
This is routinely interpreted as assurance that the actual value
(among all thinkable values)
is in this range with 95\% probability.
Although this interpretation is temptingly suggested by the terminology,
it is incorrect
--- one must not have such confidence in a confidence interval.

Rather, the situation is this:
After defining a full set of confidence intervals for quantity $F$, one
interval for each thinkable data, the confidence level of the set is its
so-called \emph{coverage}, which is the fraction of intervals that cover the
actual value, minimized over all possible $F$ values, whereby each interval is
weighted by the probability of observing the data associated with it.
Upon denoting the confidence interval for data $D$ by $\cC_{D}$, the coverage
of the set $\mathbf{C}=\{\cC_D\}$ is thus calculated in accordance with
\begin{equation}\label{eq:CC-1}
  \cov(\mathbf{C})=\min_{F}\sum_DL(D|F)\left\{
      \begin{array}{c@{\ \mbox{if}\ }l}
       1 & F\in\cC_D  \\ 0 & F\not\in\cC_D
      \end{array}\right\}\,.
\end{equation}
We emphasize that the coverage is a property of the set, not of any individual
confidence interval; the whole set is needed for associating a level of
confidence with the intervals that compose the set.

A set $\textbf{C}$ of confidence intervals with coverage
$\cov(\textbf{C})=0.95$ has this meaning:
If we repeat the experiment very often and find the respective interval
$\cC_D$ for each data $D$ obtained, then 95\% of these intervals will contain
the actual value of $F$.
Confidence intervals are a concept of frequentism where the notion of
probability refers to asymptotic relative frequencies --- the confidence
intervals are random while the actual value of $F$ is whatever it is (yet
unknown to us) and 95\% of the confidence intervals contain it.
Here we do statistics on the intervals, not on the value of $F$.
It is incorrect to infer that, for each 95\% confidence interval,
there are odds of 19:1 in favor of containing the actual value of $F$,
an individual confidence interval conveys no such information.

It is possible, as demonstrated by the example that follows below,
that the confidence interval associated with the observed
data contains the actual value of $F$ certainly, or certainly not, and that
the data tell us about this.
This can even happen for each confidence interval in a set determined by
standard optimality criteria \cite{Plante:91}
(see also Example 3.4.3 in \cite{Evans:15}).

The example just alluded to is a scenario invented by Jaynes
\cite{Jaynes:76} (see also \cite{VanderPlas:14}).
We paraphrase it as follows:
A certain process runs perfectly for duration $T$, after which failures occur
at a rate $r$, so that the probability of observing the first failure between
time $t$ and $t+\D t$ is
\begin{equation}\label{eq:CC-2}
  \D t\, r\, \Exp{-r(t-T)}\eta(t-T)\,.
\end{equation}
We cannot measure $T$ directly; instead we record first-failure times
$t_1,t_2,\dots,t_N$ when restarting the process $N$ times.
Question: What do the data $D=\{t_1,t_2,\dots,t_N\}$ tell us about $T$?

One standard frequentist approach begins with noting that the expected
first-failure time is
\begin{equation}\label{eq:CC-3}
  \mathbb{E}(t)=\int_{-\infty}^{\infty}\D t\, r\, \Exp{-r(t-T)}\eta(t-T)\,t
=T+\frac{1}{r}\,.
\end{equation}
Since the average $\tav$ of the observed failure times,
\begin{equation}\label{eq:CC-4}
  \tav=\frac{1}{N}\sum_{n=1}^N t_n\,,
\end{equation}
is an estimate for $\mathbb{E}(t)$, we are invited to use
\begin{equation}\label{eq:CC-5}
  \widehat{T}=\tav-\frac{1}{r}
\end{equation}
as the point estimator for $T$.
In many repetitions of the experiment, then, the probability of obtaining the
estimator between $\widehat{T}$ and $\widehat{T}+\D\widehat{T}$ is
\begin{eqnarray}\label{eq:CC-6}
  &&\int_T^{\infty}\D t_1\,r\,\Exp{-r(t_1-T)}
  \int_T^{\infty}\D t_2\,r\,\Exp{-r(t_2-T)}\cdots
  \nonumber\\&&\times\int_T^{\infty}\D t_N\,r\,\Exp{-r(t_N-T)}
  \delta\biggl(\widehat{T}-\tav+\frac{1}{r}\biggr)\D\widehat{T}
\nonumber\\&=&\D\widehat{T}\,f_N(\widehat{T}-T)
\end{eqnarray}
with
\begin{equation}\label{eq:CC-7}
  f_N(t)=Nr\frac{[N(rt+1)]^{N-1}}{(N-1)!}\Exp{-N(rt+1)}\eta(rt+1)\,.
\end{equation}
Accordingly, the expected value of $\widehat{T}$ is $T$,
\begin{equation}\label{eq:CC-8}
  \mathbb{E}(\widehat{T})
=\int_{-\infty}^{\infty}\D\widehat{T}\,f_N(\widehat{T}-T)\,\widehat{T}=T\,,
\end{equation}
which says that the estimator of Eq.~\eqref{CC-5} is unbiased.
It is also consistent (the more important property) since
\begin{equation}\label{eq:CC-9}
  f_N(\widehat{T}-T)\xrightarrow{N\to\infty}\delta(\widehat{T}-T)\,.
\end{equation}

Next, we consider the set $\mathbf{C}_N(t_1,t_2)$ of intervals specified by
\begin{equation}\label{eq:CC-10}
  \widehat{T}-t_1<T<\widehat{T}+t_2
\end{equation}
and establish its coverage,
\begin{eqnarray}\label{eq:CC-11}
  \cov\bigl(\mathbf{C}_N(t_1,t_2)\bigr)
&=&\min_T\int_{-\infty}^{\infty}\D\widehat{T}\,f_N(\widehat{T}-T)
\nonumber\\&&\hphantom{\min_T\int}\times
\eta(T-\widehat{T}+t_1)\eta(\widehat{T}-T+t_2)\nonumber\\
&=&\int\power{y_1}\rewop{\min\{0,y_2\}}
\D y\frac{y^{N-1}}{(N-1)!}\,\Exp{-y}
\end{eqnarray}
with ${y_1=N(rt_1+1)}$ and ${y_2=N(1-rt_2)<y_1}$.
Of the $y_1,y_2$ pairs that give a coverage of $0.95$, one would usually not
use the pairs with ${y_1=\infty}$ or ${y_2=0}$ but rather opt for the pair
that gives the shortest intervals --- the frequentist analog of the smallest
credible intervals.
These shortest intervals are obtained by the restrictions
\begin{eqnarray}
  \label{eq:CC-12}
  &&0<y_2<N-1<y_1<\infty\nonumber\\[1ex] \mbox{with}&\quad&
 y_2^{N-1}\Exp{-y_2} =y_1^{N-1}\Exp{-y_1}
\end{eqnarray}
on $y_1$ and $y_2$ in Eq.~\eqref{CC-11}.

When ${N=3}$, we have ${y_1=6.400}$ and
${y_2=0.3037}$, and the shortest confidence intervals with $95\%$ coverage are
given by
\begin{equation}\label{eq:CC-13}
  \frac{1}{3}{\left(\sum_{n=1}^3t_n-\frac{6.400}{r}\right)}<T<
  \frac{1}{3}{\left(\sum_{n=1}^3t_n-\frac{0.3037}{r}\right)}\,.
\end{equation}
There is, for instance \cite{VanderPlas:14}, the interval associated with the
data ${t_1=10/r}$, ${t_2=12/r}$, and ${t_3=15/r}$,
\begin{equation}\label{eq:CC-14}
  \frac{10.2}{r} < T <  \frac{12.2}{r}\,.
\end{equation}
Most certainly, the actual value of $T$ is \emph{not inside} this 95\%
confidence interval since $T$ must be less than the earliest observed failure
time,
\begin{equation}\label{eq:CC-15}
  T<\tmin=\min_n\{t_n\}\,,
\end{equation}
here: ${T<10/r}$.
By contrast, the 95\% confidence interval for the data ${t_1=1.9/r}$,
${t_2=2.1/r}$, and ${t_3=2.3/r}$, namely
\begin{equation}\label{eq:CC-14'}
  -\frac{0.03}{r} < T <  \frac{2.00}{r}\,,
\end{equation}
contains all values between ${T=0}$ and ${T=\tmin=1.9/r}$, so that the actual
value is \emph{certainly inside.}

These examples illustrate well what is stated above:
The interpretation ``the actual value is inside this 95\% confidence interval
with 95\% probability'' is incorrect.
Jaynes's scenario is particularly instructive because the data tell us
that the confidence interval of Eq.~\eqref{CC-14} is completely off target
and that of Eq.~\eqref{CC-14'} is equally useless.
Clearly, these 95\% confidence intervals do not answer the
question asked above: What do the data tell us about $T$?

This is not the full story, however.
The practicing frequentist can use alternative strategies for constructing
sets of shortest confidence intervals.
There is, for example, another standard method
that takes the maximum-likelihood point estimator as its starting point.
The point likelihood for observing first failures at times $t_1,t_2,\dots,t_N$
is
\begin{eqnarray}\label{eq:CC-16}
  L(D|T)&=&\prod_{n=1}^Nr\tau\,\Exp{-r(t_n-T)}\eta(t_n-T)\nonumber\\
&=&L_{\mathrm{max}}(D)\Exp{-Nr(\tmin-T)}\eta(\tmin-T)\qquad
\end{eqnarray}
where $\tau\ll1/r$ is the precision of the observations and the maximal value
\begin{equation}\label{eq:CC-17}
  L_{\mathrm{max}}(D)=L(D|T=\TML)=(r\tau)^N\Exp{-Nr(\tav-\tmin)}
\end{equation}
is obtained for the maximum-likelihood estimator $\TML=\tmin$.
In this case, $f_N(\widehat{T}-T)$ of Eqs.~\eqref{CC-6} and \eqref{CC-7} is
replaced by
\begin{equation}\label{eq:CC-23}
  \widehat{T}=\tmin:\quad
  f_N(\widehat{T}-T)=Nr\,\Exp{-Nr(\widehat{T}-T)}\,\eta(\widehat{T}-T)\,,
\end{equation}
which, not accidentally, is strikingly similar to the likelihood $L(D|T)$ in
Eq.~\eqref{CC-16} but has a completely different meaning.
Since Eq.~\eqref{CC-9} holds, this estimator is consistent, and it has a bias,
\begin{equation}\label{eq:CC-24}
\mathbb{E}(\widehat{T})
=\int_{-\infty}^{\infty}\D\widehat{T}\,f_N(\widehat{T}-T)\,\widehat{T}
=T+\frac{1}{Nr}\neq T\,,
\end{equation}
that could be removed.
The resulting shortest confidence intervals are specified by
\begin{equation}\label{eq:CC-26}
\tmin-\frac{1}{Nr}\log\frac{1}{1-\cov(\mathbf{C})}<T<\tmin\,,
\end{equation}
where $\cov(\mathbf{C})$ is the desired coverage of the set $\mathbf{C}$ thus
defined.
Here, we obtain the 95\% confidence intervals
\begin{equation}\label{eq:CC-81}
  \frac{9.0}{r} < T <  \frac{10.0}{r}\quad\mbox{and}\quad
  \frac{0.90}{r} < T <  \frac{1.90}{r}
\end{equation}
for the ${N=3}$ data that yielded the intervals in Eqs.~\eqref{CC-14} and
\eqref{CC-14'}.

While this suggests, and rather strongly so, that the confidence intervals of
this second kind are more reasonable and more useful than the previous ones,
it confronts us with the need for a criterion by which we select the
preferable set of confidence intervals among equally legitimate sets.
Chernoff offers pertinent advice for that \cite{Chernoff-quote}:
``Start out as a Bayesian thinking about it, and you'll get the right answer.
Then you can justify it whichever way you like.''

So, let us now find the corresponding SCIs of the Bayesian approach,
where probability quantifies our belief --- in
colloquial terms: Which betting odds would we accept?
For the point likelihood of Eq.~\eqref{CC-16},
the  BLI $\cI_{\lambda}$ is specified by
\begin{equation}\label{eq:CC-18}
  \max{\left\{0,\tmin-\frac{1}{Nr}\log\frac{1}{\lambda}\right\}}<T<\tmin\,.
\end{equation}
Jaynes recommends a flat prior in such applications --- unless we have specific
prior information about $T$, that is --- but, without a restriction on the
permissible $T$ values, that would be an improper prior here.
Instead we use $\D T\,\kappa\,\Exp{-\kappa T}\eta(T)$ for the prior element
and enforce ``flatness'' by taking the limit of ${\kappa\to0}$ eventually.
Then, the likelihood for the observed data is
\begin{eqnarray}\label{eq:CC-19}
  L(D)&=&\int_0^{\infty}\D T\,\kappa\,\Exp{-\kappa t}\,L(D|T)
\nonumber\\ &=&L_{\mathrm{max}}(D)\frac{\kappa}{Nr-\kappa}
{\left(\Exp{-\kappa\tmin}-\Exp{-Nr\tmin}\right)}\qquad
\end{eqnarray}
and the credibility of $\cI_{\lambda}$ is
\begin{eqnarray}\label{eq:CC-20}
  c_{\lambda}&=&\int_0^{\infty}\D T\,\kappa\,\Exp{-\kappa t}\,
         \frac{L(D|T)}{L(D)}
    \eta{\left(T-\tmin+\frac{1}{Nr}\log\frac{1}{\lambda}\right)}
\nonumber\\
&\makebox[0pt][l]{$\ds\xrightarrow{\kappa\to0}\min{\left\{1,\frac{1-\lambda}
{1-\Exp{-Nr\tmin}}\right\}}$}&
\end{eqnarray}
after taking the ${\kappa\to0}$ limit.
We so arrive at
\begin{equation}\label{eq:CC-21}
  \tmin-\frac{1}{Nr}\log\frac{1}{(1-c)+\Exp{-Nr\tmin}c}<T<\tmin
\end{equation}
for the SCI with pre-chosen credibility $c$.
For example, the SCIs for ${c=0.95}$ that corresponds to the confidence
intervals in Eqs.~\eqref{CC-14} and \eqref{CC-14'}, and also to the confidence
intervals in Eq.~\eqref{CC-81}, are
\begin{equation}\label{eq:CC-22}
  \frac{9.0}{r} < T <  \frac{10.0}{r}\quad\mbox{and}\quad
  \frac{0.92}{r} < T <  \frac{1.90}{r}\,.
\end{equation}
These really \emph{are} useful answers to the question of what do the data tell
us about $T$: The actual value is in the respective range with 95\% probability.

Regarding the choice between the set of confidence intervals of the first and
the second kind --- associated with the point estimators ${\widehat{T}=\tav}$
and  ${\widehat{T}=\tmin}$, respectively --- Chernoff's strategy clearly
favors the second kind.
Except for the possibility of getting a negative value for the lower bound,
the confidence intervals of Eq.~\eqref{CC-26} are the BLIs of
Eq.~\eqref{CC-18} for
$\lambda=1-\cov(\textbf{C})$, and they are virtually identical with the SCIs of
Eq.~\eqref{CC-21} --- usually the term $\Exp{-Nr\tmin}c$ is negligibly small
there.
Yet, these confidence intervals retain their frequentist meaning.

Such a coincidence of confidence intervals and credible intervals is also
possible under other circumstances, and this observation led Jaynes to the
verdict that
``confidence intervals are
satisfactory as inferences \emph{only} in those special cases where they
happen to agree with Bayesian intervals after all'' (Jaynes's emphasis, see
p.~674 in \cite{Jaynes:03}).
That is: One can get away with misinterpreting the confidence intervals
as credible intervals for an unspecified prior.

In the context of the example we are using, the coincidence occurs as a
consequence of two ingredients: (i) We are guided by the Bayesian reasoning
when choosing the set of confidence intervals;
(ii) we are employing the flat prior when determining the SCIs.
The coincidence does not happen when (i) another strategy is used for the
construction of the set of confidence intervals, or (ii) for another prior, as
we would use it if we had genuine prior information about $T$;
the coincidence could still occur when $N$ is large but hardly for ${N=3}$.

In way of summary, the fundamental difference between the
confidence intervals of Eqs.~\eqref{CC-14} and \eqref{CC-14'}, or those of
Eq.~\eqref{CC-81}, and the credible intervals of Eq.~\eqref{CC-22},
which refer to the same data, is this:
We judge the quality (= confidence level = coverage) of the confidence
interval $\cC_D$ by the company it keeps (= the full set $\mathbf{C}=\{\cC_D\}$),
whereas the credible interval is judged on its own merits (= credibility).
It is worth repeating here that the two types of intervals tell us about very
different things:
Confidence intervals are about statistics on the data; credible intervals are
about statistics on the quantity of interest.
If one wishes, as we do, to draw reliable conclusions from the data of a
single run, one should  use the Bayesian credible interval and not the
frequentist confidence interval.

What about many runs?
If we take, say, one hundred measurements of three first-failure times, we can
find the one hundred shortest 95\% confidence intervals of either kind
and base our conclusions on the properties of this set.
Alternatively, we can combine the data and regard them as three
hundred first-failure times of a single run and so arrive at a SCI with a size
that is one-hundredth of each SCI for three first-failure times.

Misconceptions such as ``confidence regions have a natural Bayesian
interpretation as regions which are credible for any prior''
\cite{NJPexpert:16}, as widespread as they may be, arise when the fundamental
difference in meaning between confidence intervals and credible intervals is
not appreciated.
While, obviously, one can compute the credibility of any region for any
prior, there is no point in this ``natural Bayesian interpretation''
for a set of confidence regions; the credibility thus found for
a particular confidence region has no universal relation to the coverage of
the set.
It is much more sensible to determine the SCRs for the data actually observed.

On the other hand, it can be very useful to pay attention to the corresponding
credible regions when constructing a set of confidence regions.
In the context of QSE, this Chernoff-type strategy is employed by Christandl
and Renner \cite{Christandl+1:12}
who take a set of credible regions and enlarge all of them to
produce a set of confidence regions; see also \cite{Blume-Kohout:12}.

Another instance where a frequentist approach benefits from Bayesian methods
is the marginalization of nuisance parameters in \cite{Faist+1:16} where a
MC integration employs a flat prior, apparently chosen because it is easy
to implement.
The histograms thus produced --- they report differences of
$\prefD(F)$ in Eq.~\eqref{6-2} between neighboring $F$ values, just like the
binned probabilities in Fig.~\figref{CHSH-histo} --- depend on
the prior, and so do the confidence intervals inferred from the histograms.

There is also a rather common misconception about the subjectivity or
objectivity of the two methods.
The frequentist confidence regions are regarded as objective, in contrast to
the subjective Bayesian credible regions.
The subjective nature of the credible regions originates in the necessity of a
prior, privately chosen by the scientist who evaluates the data and properly
accounts for her prior knowledge.
No prior is needed for the confidence regions, they are completely determined
by the data --- or so it seems.
In fact, the choice between different sets of confidence regions is equally
private and subjective; in the example above, it is the choice between the
confidence intervals of Eqs.~\eqref{CC-10}--\eqref{CC-12}, those of
Eq.~\eqref{CC-26}, and yet other legitimate constructions which, perhaps, pay
attention to prior knowledge.
Clearly, either approach has unavoidable subjective ingredients, and this
requires that we state, completely and precisely, how the data are processed;
see Sec.~1.5.2 in \cite{Evans:15} for further pertinent remarks.

\section{Prior-content function $P_0(\CHSH)$ near ${\CHSH=\pm\sqrt{8}}$}
\label{sec:appA}
In this appendix, we consider the sizes of the regions with
${\chsh(p)\gtrsim-\sqrt{8}}$ and ${\chsh(p)\lesssim\sqrt{8}}$.
It is our objective to justify the power law
stated in Eq.~\eqref{8-13} and so motivate the approximation in Eq.~\eqref{8-14}.

We denote the kets of the maximally entangled states with
${\chsh=\pm\sqrt{8}}$ by
$\ket{\pm}$, that is
\begin{equation}\label{eq:A-1}
  \ket{+}=\frac{\ket{\up\dn}-\ket{\dn\up}}{\sqrt{2}}
  \quad\mbox{and}\quad
  \ket{-}=\frac{\ket{\up\up}+\ket{\dn\dn}}{\sqrt{2}}
\,,
\end{equation}
where ${\ket{\up\dn}=\ket{\up}\otimes\ket{\dn}}$, for example, has
${\sigma_z=1}$ for the first qubit and ${\sigma_z=-1}$ for the second.
Since
\begin{equation}\label{eq:A-2}
  \sigma_x\otimes\identity\ket{\pm}=\mp\identity\otimes\sigma_x\ket{\pm}
\quad\mbox{and}\quad
  \sigma_z\otimes\identity\ket{\pm}=\mp\identity\otimes\sigma_z\ket{\pm}\,,
\end{equation}
we have [recall Eq.~\eqref{8-5}]
\begin{eqnarray}\label{eq:A-3}
  \Pi_k^{\ }\ket{\pm}&=&\Pi_{[jj']}^{\ }\ket{\pm}
=\Pi_j^{(1)}\otimes\Pi^{(2)}_{j'}\ket{\pm}\nonumber\\&=&
\frac{1}{9}(1+\vec{t}_j\cdot\bfsym{\sigma})\otimes
(1-\vec{t}_{j'}\cdot\bfsym{\sigma})\ket{\pm}\nonumber\\&=&
\frac{1}{9}(1+\vec{t}_j\cdot\bfsym{\sigma})
            (1\pm\vec{t}_{j'}\cdot\bfsym{\sigma})\otimes\identity\ket{\pm}
\nonumber\\&=&
\frac{1}{9}\Bigl[(1\pm\vec{t}_j\cdot\vec{t}_{j'})\identity\nonumber\\
&&\phantom{\frac{1}{9}\Bigl[}
+(\vec{t}_j\pm\vec{t}_{j'}\pm\I\vec{t}_j\bfsym{\times}\vec{t}_{j'})\cdot
\bfsym{\sigma}\Bigr]
\otimes\identity\ket{\pm}\,,\qquad
\end{eqnarray}
where
\begin{equation}\label{eq:A-4}
  \vec{t}_1=\vec{e}_z\,,\quad
  \vec{t}_2=\frac{\sqrt{3}}{2}\vec{e}_x-\frac{1}{2}\vec{e}_z\,,\quad
  \vec{t}_3=-\frac{\sqrt{3}}{2}\vec{e}_x-\frac{1}{2}\vec{e}_z
\end{equation}
are the three unit vectors of the trine.

States in an $\epsilon$-vicinity of $\ketbra{\pm}$ are of the form
\begin{eqnarray}\label{eq:A-5}
  \rho_{\epsilon}^{\ }&=&
\frac{\bigl(\ketbra{\pm}+\epsilon A^{\dagger}\bigr)
      \bigl(\ketbra{\pm}+\epsilon A\bigr)}
     {1+\epsilon\bra{\pm}(A^\dagger+A)\ket{\pm}+\epsilon^2\tr{A^\dagger A}}
\nonumber\\&=&
\ketbra{\pm}+\epsilon\,(A_{\pm}^\dagger+A_{\pm}^{\ })+O(\epsilon^2)\,,
\end{eqnarray}
where $A$ is any two-qubit operator and
\begin{equation}\label{eq:A-6}
  A_{\pm}^{\ }=\ketbra{\pm}A\bigl(\identity-\ketbra{\pm}\bigr)
\end{equation}
is a traceless rank-1 operator with the properties
\begin{equation}\label{eq:A-6a}
  A_{\pm}^{\ }\ket{\pm}=0
\quad\mbox{and}\quad
\ketbra{\pm}A_{\pm}^{\ }=A_{\pm}^{\ }\,.
\end{equation}
The TAT probabilities are
\begin{eqnarray}\label{eq:A-7}
  p_k&=&\tr{\rho_{\epsilon}\Pi_k}\nonumber\\&=&\bra{\pm}\Pi_k\ket{\pm}
        +\epsilon\,\tr{(A_{\pm}^\dagger+A_{\pm}^{\ })\Pi_k}+O(\epsilon^2)
\nonumber\\&=&
\frac{1}{9}\bigl(1\pm\vec{t}_j\cdot\vec{t}_{j'}\bigr)
+\epsilon\Bigl[(\vec{t}_j\pm\vec{t}_{j'})\cdot\bfsym{\alpha}_{\pm}^{\ }
\mp(\vec{t}_j\bfsym{\times}\vec{t}_{j'})\cdot\bfsym{\beta}_{\pm}^{\ }\Bigr]
\nonumber\\
&&+O(\epsilon^2)
\end{eqnarray}
with the real vectors $\bfsym{\alpha}_{\pm}^{\ }$ and
$\bfsym{\beta}_{\pm}^{\ }$ given by
\begin{equation}\label{eq:A-8}
  \frac{2}{9}\tr{\bfsym{\sigma}\otimes\identity A_{\pm}^{\ }}
  =\bfsym{\alpha}_{\pm}^{\ } +\I \bfsym{\beta}_{\pm}^{\ }\,.
\end{equation}
Owing to the trine geometry, the $x$ and $z$ components of
$\bfsym{\alpha}_{\pm}^{\ }$ and the $y$ component of
$\bfsym{\beta}_{\pm}^{\ }$ matter, but the other three components do not.
In the eight-dimensional probability space, then, we have increments
${\propto\epsilon}$ in three directions only, and increments
${\propto\epsilon^2}$ in the other five directions.
For the primitive prior, therefore, the size of the $\epsilon$-vicinity is
${\propto\epsilon^{3\times1+5\times2}=\epsilon^{13}}$.

The sum of probabilities in Eq.~\eqref{8-6} is
\begin{equation}\label{eq:A-9}
  p_1+p_5+p_9=p_{[11]}+p_{[22]}+p_{[33]}=\frac{1}{3}(1\pm1)+O(\epsilon^2)\,,
\end{equation}
so that ${\CHSH=\pm\sqrt{8}\,[1-O(\epsilon^2)]}$ or
${\sqrt{8}-\bfsym{|}\CHSH\bfsym{|}\propto\epsilon^2}$.
Accordingly, we infer that
\begin{equation}\label{eq:A-10}
  P_0(\CHSH)\propto{\left(\sqrt{8}+\CHSH\right)}^\frac{13}{2}
  \quad\mbox{near ${\CHSH=-\sqrt{8}}$}
\end{equation}
and
\begin{equation}\label{eq:A-11}
  1-P_0(\CHSH)\propto{\left(\sqrt{8}-\CHSH\right)}^\frac{13}{2}
  \quad\mbox{near ${\CHSH=\sqrt{8}}$}\,,
\end{equation}
which imply Eq.~\eqref{8-13}.

\section{Prior-content function $P_0(\CHSHopt)$ near
${\CHSHopt=0}$ and ${\CHSHopt=\sqrt{8}}$
}\label{sec:appB}
In this appendix, we consider the sizes of the regions with
${\CHSHopt\gtrsim0}$ and ${\CHSHopt\lesssim\sqrt{8}}$.
We wish to establish the $\CHSHopt$ analogs of Eqs.~\eqref{8-13} and \eqref{8-14}.

In the context of $P_0(\CHSHopt)$, it is expedient to switch from the
nine TAT probabilities $p_1,p_2,\dots,p_9$ to the expectation values of the
eight single-qubit and two-qubit observables that are linearly related to the
probabilities,
\begin{widetext}
\begin{eqnarray}\label{eq:B-1}
 {\left(\begin{array}{@{}ccc@{}}
  p_1 & p_2 & p_3 \\
  p_4 & p_5 & p_6 \\
  p_7 & p_8 & p_9
  \end{array}\right)}
\begin{array}{c}
  \mbox{\footnotesize{}linear}\\[-1.5ex]
  \longleftarrow\hspace*{-0.5em}\frac{\quad}{}\hspace*{-0.5em}\longrightarrow
\\[-1.5ex]\mbox{\footnotesize{}relation}
\end{array}
 {\left[\begin{array}{@{}c|cc@{}}
      & \expect{\identity\otimes\sigma_x} &
        \expect{\identity\otimes\sigma_z} \\ \hline
  \expect{\sigma_x\otimes\identity} & \expect{\sigma_x\otimes\sigma_x}
          & \expect{\sigma_x\otimes\sigma_z} \\
  \expect{\sigma_z\otimes\identity} & \expect{\sigma_z\otimes\sigma_x}
          & \expect{\sigma_z\otimes\sigma_z}
  \end{array}\right]}\equiv
 {\left[\begin{array}{@{}c|cc@{}}
      & x_3 & x_4 \\ \hline
  x_1 & y_1 & y_2 \\
  x_2 & y_3 & y_4
  \end{array}\right]}.
\end{eqnarray}
The Jacobian matrix associated with the linear relation does not depend on the
probabilities and, therefore, we have
\begin{equation}\label{eq:B-2}
  (\D\rho)=(\D p)=(\D x)\,(\D y)\,w_{\mathrm{cstr}}(x,y)
\end{equation}
for the primitive prior, where
$(\D x)=\D x_1\,\D x_2\,\D x_3\,\D x_4$ and
$(\D y)=\D y_1\,\D y_2\,\D y_3\,\D y_4$,
and $w_{\mathrm{cstr}}(x,y)$ equals a normalization factor for
permissible values of ${x=(x_1,x_2,x_3,x_4)}$ and ${y=(y_1,y_2,y_3,y_4)}$,
whereas ${w_{\mathrm{cstr}}(x,y)=0}$ for unphysical values.
Thereby, the permissible values of $x$ and $y$ are those for which one can
find $q$ in the range ${-1\leq q\leq1}$ such that \cite{note11}
\begin{equation}\label{eq:B-4}
  {\left(\begin{array}{@{}c@{\,}c@{\,}c@{\,}c@{}}
  1+x_1+x_3+y_1  &     x_2+y_3      & x_4+y_2  & y_4-q \\
     x_2+y_3     &  1-x_1+x_3-y_1  & y_4+q  & x_4-y_2 \\
     x_4+y_2     &     y_4+q      &  1+x_1-x_3-y_1  &  x_2-y_3 \\
     y_4-q       &    x_4-y_2     &   x_2-y_3  &  1-x_1-x_3+y_1
  \end{array}\right)}\geq0\,.
\end{equation}
While the implied explicit conditions on $x$ and $y$ are rather involved,
the special cases of interest here --- namely ${x=0}$ and ${y=0}$,
respectively --- are quite transparent.
We have
\begin{equation}\label{eq:B-5}
  w_{\mathrm{cstr}}(x,0)=0 \enskip
\mbox{unless $\ds{\left(x_1^2+x_2^2\right)}^{\frac{1}{2}}
                +{\left(x_3^2+x_4^2\right)}^{\frac{1}{2}}\leq1$}
\end{equation}
and
\begin{eqnarray}\label{eq:B-6}
  w_{\mathrm{cstr}}(0,y)=0\enskip
&&\mbox{unless the two characteristic values}\nonumber\\
&&\mbox{of
${\left(\begin{array}{@{}cc@{}}y_1 & y_2 \\ y_3 & y_4\end{array}\right)}$
are ${\leq1}$.}
\end{eqnarray}
The sum of the squares of these characteristic values is
${y_1^2+y_2^2+y_3^2+y_4^2}$; it determines the value of $\chshopt(p)$,
\begin{equation}\label{eq:B-7}
  \chshopt=2{\left(y_1^2+y_2^2+y_3^2+y_4^2\right)}^{\frac{1}{2}}\,.
\end{equation}
\end{widetext}

\subsection{The vicinity of ${\CHSHopt=0}$}\label{sec:appB-1}
We obtain ${\chshopt=0}$ for ${y=0}$ and
\begin{eqnarray}\label{eq:B-8}
  {\left(\begin{array}{@{}c@{}} x_1 \\x_2\end{array}\right)}&=&
{\left(\begin{array}{@{}cc@{}} \cos\varphi_1 & -\sin\varphi_1 \\
                              \sin\varphi_1 & \cos\varphi_1\end{array}\right)}
{\left(\begin{array}{@{}c@{}} r_1 \\ 0 \end{array}\right)}
={\left(\begin{array}{@{}c@{}} r_1\cos\varphi_1 \\ r_1\sin\varphi_1
        \end{array}\right)}\,,
\nonumber\\
 {\left(\begin{array}{@{}cc@{}} x_3 & x_4\end{array}\right)}&=&
  {\left(\begin{array}{@{}cc@{}} r_2 &  0 \end{array}\right)}
{\left(\begin{array}{@{}cc@{}} \cos\varphi_2 & \sin\varphi_2 \\
                           -\sin\varphi_2 & \cos\varphi_2\end{array}\right)}
\nonumber\\
&=&
{\left(\begin{array}{@{}cc@{}} r_2\cos\varphi_2 & r_2\sin\varphi_2
\end{array}\right)}\,,
\nonumber\\
(\D x)&=&\D r_1\,r_1\,\D\varphi_1\,\D r_2\,r_2\,\D\varphi_2
\end{eqnarray}
\par\noindent{}
with ${0\leq r_1\leq1-r_2\leq1}$ and ${0\leq\varphi_1,\varphi_2\leq2\pi}$.
These $x$ values make up a four-dimen\-sional volume
\begin{equation}\label{eq:B-9}
  \int(\D x)=(2\pi)^2\int_0^1\D r_1\,r_1\int_0^{1-r_1}\D r_2\,r_2
=\frac{\pi^2}{6}
\end{equation}
but, since there is no volume in the four-dimensional $y$ space, the set of
probabilities with ${\chshopt=0}$ has no eight-dimensional volume --- it has no
size.

The generic state in this set has ${r_1+r_2<1}$ and full rank.
A finite, if small, four-dimensional ball is then available for the $y$
values.
All $y$ values on the three-dimensional surface of the ball have the same
value of $\chshopt$, equal to the diameter of the ball.
The volume of the ball is proportional to $\chshopt^4$ and,
therefore, we have
\begin{equation}\label{eq:B-10}
  P_0(\CHSHopt)\propto\CHSHopt^4\quad\mbox{for}\quad 0\lesssim\CHSHopt\ll1\,.
\end{equation}

\begin{widetext}
\subsection{The vicinity of ${\CHSHopt=\sqrt{8}}$}\label{sec:appB-2}
We reach ${\CHSHopt=\sqrt{8}}$ for all maximally entangled states with
${\expect{\sigma_y\otimes\sigma_y}^2=1}$.
Then, ${x=0}$ and both characteristic values of the ${2\times2}$ matrix in
Eq.~\eqref{B-6} are maximal.
More generally, when ${x=0}$, the permissible $y$ values are
\begin{eqnarray}\label{eq:B-11}
{\left(\begin{array}{@{}cc@{}}y_1 & y_2 \\ y_3 & y_4\end{array}\right)}&=&
{\left(\begin{array}{@{}cc@{}}\cos\phi_1 & -\sin\phi_1 \\
                              \sin\phi_1 &  \cos\phi_1\end{array}\right)}
{\left(\begin{array}{@{}cc@{}}\vartheta_1 & 0 \\ 0 &
          \vartheta_2\end{array}\right)}
{\left(\begin{array}{@{}cc@{}} \cos\phi_2 & \sin\phi_2 \\
                              -\sin\phi_2 & \cos\phi_2\end{array}\right)}
\\\hspace*{-1.5em}&=&
{\left(\begin{array}{@{}cc@{}}
  \vartheta_1\cos\phi_1\cos\phi_2 +\vartheta_2\sin\phi_1\sin\phi_2 &
  \vartheta_1\cos\phi_1\sin\phi_2 -\vartheta_2\sin\phi_1\cos\phi_2 \\
  \vartheta_1\sin\phi_1\cos\phi_2 -\vartheta_2\cos\phi_1\sin\phi_2 &
  \vartheta_1\sin\phi_1\sin\phi_2 +\vartheta_2\cos\phi_1\cos\phi_2
\end{array}\right)} \nonumber
\end{eqnarray}
\end{widetext}
with ${0\leq \vartheta_1\leq1}$, ${-1\leq \vartheta_2\leq1}$,
${0\leq\phi_1,\phi_2\leq2\pi}$, where $\vartheta_1$ and
$\bfsym{|}\vartheta_2\bfsym{|}$ are the characteristic values.
The determinant $\vartheta_1\vartheta_2$ can be positive or negative; we avoid
double coverage by restricting $\vartheta_1$ to positive values while letting
$\phi_1$ and $\phi_2$ range over a full $2\pi$ period.

The Jacobian factor in
\begin{equation}\label{eq:B-12}
  (\D y)=\D \vartheta_1\,\D \vartheta_2\,\D\phi_1\,\D\phi_2\,
         \bfsym{|}\vartheta_1^2-\vartheta_2^2\bfsym{|}
\end{equation}
vanishes when ${\vartheta_1=\bfsym{|}\vartheta_2\bfsym{|}=1}$ and
$\CHSHopt=2{\left(\vartheta_1^2+\vartheta_2^2\right)}^{\frac{1}{2}}=\sqrt{8}$.
Therefore, there is no nonzero four-dimensional volume in the $y$ space for
${\CHSHopt=\sqrt{8}}$.
More specifically, the $y$-space volume for
${\vartheta_1^2+\vartheta_2^2>\frac{1}{4}\CHSHopt^2}$ is
\begin{eqnarray}\label{eq:B-13}
  &&(2\pi)^2\int_0^1\D \vartheta_1\int_{-1}^1\D \vartheta_2\,
\bfsym{|}\vartheta_1^2-\vartheta_2^2\bfsym{|}
\,\eta\Bigl(4{\left(\vartheta_1^2+\vartheta_2^2\right)}-\CHSHopt^2\Bigr)
\nonumber\\&=&(2\pi)^2{\left[\frac{2}{3}-\frac{1}{32}\CHSHopt^4
+\frac{1}{6}{\left(\CHSHopt^2-4\right)}^\frac{3}{2}\,
\eta{\left(\CHSHopt^2-4\right)}
\right]}
\nonumber\\&=&\frac{\sqrt{8}\,\pi^2}{3}
{\left(\sqrt{8}-\CHSHopt\right)}^3
+O{\left({\left(\sqrt{8}-\CHSHopt\right)}^4\right)}\nonumber\\
&&\mbox{for}\quad\CHSHopt\lesssim\sqrt{8}\,.
\end{eqnarray}

With respect to the corresponding $x$-space volume, we note that the maximally
entangled states with
\begin{align}\label{eq:B-14}
&{\left(\begin{array}{@{}cc@{}}y_1 & y_2 \\ y_3 & y_4\end{array}\right)}=
{\left(\begin{array}{@{}cc@{}}
\cos(\phi_1-\phi_2) & \sin(\phi_1-\phi_2) \\
-\sin(\phi_1-\phi_2) &\cos(\phi_1-\phi_2)
\end{array}\right)}\nonumber\\
\mbox{or}\quad&
{\left(\begin{array}{@{}cc@{}}y_1 & y_2 \\ y_3 & y_4\end{array}\right)}=
{\left(\begin{array}{@{}cc@{}}
\cos(\phi_1+\phi_2) & \sin(\phi_1+\phi_2) \\
\sin(\phi_1+\phi_2) &-\cos(\phi_1+\phi_2)
\end{array}\right)}
\end{align}
are equivalent because local unitary transformations turn them into each
other.
It is, therefore, sufficient to consider an $\epsilon$-vicinity of one such
state, for which we take that with ${y_1=y_4=-1}$ and ${y_2=y_3=0}$.
This is $\ketbra{+}$ of Eq.~\eqref{A-1}, with $\rho_{\epsilon}$ in Eq.~\eqref{A-5}.

As a consequence of Eq.~\eqref{A-2}, we have
\begin{eqnarray}\label{eq:B-15}
&&  x_1+x_3\propto\epsilon^2\,,\enskip
  x_1-x_3\propto\epsilon\nonumber\\\mbox{and}&\quad&
  x_2+x_4\propto\epsilon^2\,,\enskip
  x_2-x_4\propto\epsilon\,,
\end{eqnarray}
so that the $x$-space volume is proportional to $\epsilon^6$.
Since we know from \eqref{A-11} that ${\sqrt{8}-\CHSHopt\propto\epsilon^2}$, it
follows that the $x$-space volume is proportional to
${\left(\sqrt{8}-\CHSHopt\right)}^3$.
Together with the $y$-space volume in \eqref{B-13}, we so find that
\begin{equation}\label{eq:B-16}
  1-P_0(\CHSHopt)\propto{\left(\sqrt{8}-\CHSHopt\right)}^6
\quad\mbox{for}\quad 0\lesssim\sqrt{8}-\CHSHopt\ll1\,.
\end{equation}

\subsection{Analog of \eqref{8-14} and \eqref{8-15} for $P_0(\CHSHopt)$}
Just like Eq.~\eqref{8-13} suggests the approximation Eq.~\eqref{8-14} for
$P_0(\CHSH)$, the power laws for $P_0(\CHSHopt)$ near ${\CHSHopt=0}$ and
${\CHSHopt=\sqrt{8}}$ in \eqref{B-10} and \eqref{B-16}, respectively, invite
the approximation
\begin{equation}\label{eq:B-17}\qquad
  P_0(\CHSHopt)\simeq P_0^{(0)}(\CHSHopt)=\sum_l w_l B_{\alpha_l,\beta_l}(\CHSHopt)
\end{equation}
with $\ds\sum_lw_l=1$ and
\begin{eqnarray}\label{eq:B-18}
  B_{\alpha,\beta}(\CHSHopt)&=&{\left(\frac{1}{8}\right)}^{\frac{1}{2}(\alpha+\beta+1)}
\frac{(\alpha+\beta+1)!}{\alpha!\;\beta!}\nonumber\\&&\times
\int_0^{\CHSHopt}\D x\,x^{\alpha}(\sqrt{8}-x)^{\beta}\,.
\end{eqnarray}
One of the powers $\alpha_l$ is equal to $3$ and one of the $\beta_l$s is
equal to $5$, and the other ones are larger.
For the sample of 500\,000 sets of probabilities that generated the red
$\CHSHopt$ histograms in Fig.~\figref{CHSH-histo}(a),
a fit with a mean squared error of $4.2\times10^{-4}$ is achieved by a
five-term approximation with these parameter values:
\begin{equation}\label{eq:B-19}
  \begin{array}{@{}crrr@{}}
    l & \multicolumn{1}{c}{w_l}& \multicolumn{1}{c}{\alpha_l}
      & \multicolumn{1}{c}{\beta_l}\\ \hline
    1 & 0.2187 & \multicolumn{1}{c}{3} & 5.2467\\
    2 & 0.2469 & 5.2238 & \multicolumn{1}{c}{5} \\
    3 & 0.3153 & 14.1703 & 11.7922\\
    4 & 0.2478 & 7.9878 & 11.8061 \\
    5 & -0.0287 & 37.5270 & 15.7518
  \end{array}
\end{equation}
There are 12 fitting parameters here.
The black curve to that histogram shows the corresponding approximation for
$\ds W_0(\CHSHopt)=\frac{\D}{\D\CHSHopt}P_0(\CHSHopt)$.

\newcommand{\prior}[1]{\par\noindent\makebox[\columnwidth][r]{%
\begin{tabular}{@{}r@{$\enskip=\enskip$}p{148pt}@{\quad\enskip}}%
#1 \end{tabular}}}

\section{List of prior densities}\label{sec:appC}
The various prior densities introduced in
Secs.~\ref{sec:stage}--\ref{sec:prior} are
\prior{$w_0(p)$ & probability-space prior density in Eq.~\eqref{2-1};}
\prior{$W_0(F)$ & prior density for property value $F$ in Eq.~\eqref{3-4};}
\prior{$u_F(p)$ & prior density on an iso-$F$ hypersurface in Eq.~\eqref{4-0a};}
\prior{$\wref(p)$ & reference prior density in Eq.~\eqref{4-0e};}
\prior{$w_{\mathrm{primitive}}(p)$ & primitive prior of Eq.~\eqref{4-6};}
\prior{$w_{\mathrm{Jeffreys}}(p)$ & Jeffreys prior of Eq.~\eqref{4-7}.\\[1ex]}
There is also the probability-space factor $w_{\mathrm{cstr}}(p)$ in
Eq.~\eqref{2-2} that accounts for the constraints.

If we choose $w_0(p)$ to our liking, then $W_0(F)$ and $u_F(p)$ are determined
by Eqs.~\eqref{3-4} and \eqref{4-0a}, respectively.
Alternatively, we can freely choose $W_0(F)$ and either $u_F(p)$ or
$\wref(p)\propto u_{f(p)}(p)$, and then obtain $w_0(p)$ from Eq.~\eqref{4-3} or
\eqref{4-0d}.
For given $u_F(p)$, the $F$-likelihood $L(D|F)$ does not depend on $W_0(F)$.

\newcommand{\acronym}[1]{\par\noindent\makebox[\columnwidth][r]{%
\begin{tabular}{@{}p{28pt}@{\qquad}p{187pt}@{}}%
#1 \end{tabular}}}

\section{List of acronyms}\label{sec:appD}
\acronym{BLI & bounded-likelihood interval}
\acronym{BLR & bounded-likelihood region}
\acronym{CHSH& Clauser-Horne-Shimony-Holt}
\acronym{CPU & central processing unit}
\acronym{DSPE& direct state-property estimation}
\acronym{ISPE& indirect state-property estimation}
\acronym{MC  & Monte Carlo}
\acronym{MLI & maximum-likelihood interval}
\acronym{POM & probability-operator measurement}
\acronym{QSE & quantum state estimation}
\acronym{SCI & smallest credible interval}
\acronym{SCR & smallest credible region}
\acronym{SPE & state-property estimation}
\acronym{TAT & trine-antitrine}


\begin{thebibliography}{00}

\bibitem{LNP649}
M.~Paris and J.~\v{R}eh\'a\v{c}ek (eds.), \textit{Quantum State Estimation},
Lecture Notes in Physics, vol.~649 (Springer-Verlag, Heidelberg, 2004).

\bibitem{Ekert+5:02}
A. K. Ekert, C. M. Alves, D. K. L. Oi, M. Horodecki, P. Horodecki, and
L. C. Kwek,
\prl~\textbf{88}, 217901 (2002).

\bibitem{Horodecki+1:02}
P. Horodecki and A. K. Ekert,
\prl~\textbf{89}, 127902 (2002).

\bibitem{Bovino+5:05}
F. A. Bovino, G. Castagnoli, A. K. Ekert, P. Horodecki, C. M. Alves,
and A. V. Sergienko,
\prl~\textbf{95}, 240407 (2005).

\bibitem{BK+2:10}
R. Blume-Kohout, J. O. S. Yin, and S. J. van Enk,
\prl~\textbf{105}, 170501 (2010).

\bibitem{Somma+2:06}
R. D. Somma, J. Chiaverini, and D. J. Berkeland,
\pra~\textbf{74}, 052302 (2006).

\bibitem{Guhne+3:07}
O. G\"{u}hne, C.-Y. Lu, W.-B. Gao, and J.-W. Pan,
\pra~\textbf{76}, 030305(R) (2007).

\bibitem{Flammia+1:11}
S. T. Flammia and Y.-K. Liu,
\prl~\textbf{106}, 230501 (2011).

\bibitem{Walborn+4:06}
S. P. Walborn, P. H. Souto Ribeiro, L. Davidovich, F. Mintert, and
A. Buchleitner,
Nature \textbf{440}, 1022 (2006).

\bibitem{Paris:09}
M. G. A. Paris,
Int.\ J. Quant.\ Inf.~\textbf{7} (Supplement), 125 (2009).

\bibitem{Shang+4:13}
J. Shang, H. K. Ng, A. Sehrawat, X. Li, and B.-G. Eng\-lert,
New J.~Phys.~\textbf{15}, 123026 (2013).
Note this erratum: The second arrow in equation (21) should be an equal sign.

\bibitem{Faist+1:16}
P. Faist and R. Renner,
\prl\ \textbf{117}, 010404 (2016).

\bibitem{Shang+4:15}
J. Shang, Y.-L. Seah, H. K. Ng, D. J. Nott, and B.-G. Englert,
New J.~Phys.~\textbf{17}, 043017 (2015).

\bibitem{Seah+4:15}
Y.-L. Seah, J. Shang, H. K. Ng, D. J. Nott, and B.-G. Englert,
New J.~Phys.~\textbf{17}, 043018 (2015).

\bibitem{Bos:02}
C. S. Bos,
\textit{A Comparison of Marginal Likelihood Computation Methods\/},
pp.~111--116 in
\textit{Compstat: Proceedings in Computational Statistics\/},
(Heidelberg: Physica-Verlag HD, 2002), edited by
W. H\"ardle and B. R\"onz.

\bibitem{Evans:15}
M. Evans, \textit{Measuring Statistical Evidence Using Relative Belief\/},
Monographs on Statistics and Applied Probability, vol.~144
(CRC Press, Boca Raton, 2015).

\bibitem{Rehacek+6:15}
J. \v{R}eh\'a\v{c}ek, Z. Hradil, Y. S. Teo, L. L. S\'anchez-Soto, H. K. Ng,
J. H. Chai, and B.-G. Englert,
\pra~\textbf{92}. 052303 (2015).

\bibitem{Jeffreys:46}
H. Jeffreys,
Proc.~R.~Soc.~Lond.~A~\textbf{186}, 453 (1946).

\bibitem{Kass+1:96}
R. E. Kass and L. Wasserman,
J.~Am.~Stat.~Assoc.~\textbf{91}, 1343 (1996).

\bibitem{Schwemmer+6:15}
C. Schwemmer, L. Knips, D. Richart, H. Weinfurter, T. Moroder,
M. Kleinmann, and O. G\"uhne,
\prl~\textbf{114}, 080403 (2015).

\bibitem{note6}
For consistency with Eq.~\eqref{3-4},
the prior density $W_0(F)$ should be positive everywhere,
except perhaps at few isolated values of $F$.

\bibitem{QSampling}
\url{http://www.quantumlah.org/publications/software/QSampling/}

\bibitem{Rehacek+1:04}
J. \v{R}eh\'a\v{c}ek, B.-G. Englert, and D. Kaszlikowski,
\pra~\textbf{70}, 052321 (2004).

\bibitem{Clauser+3:69}
J. F. Clauser, M. A. Horne, A. Shimony, and R. A. Holt,
\prl~\textbf{23}, 880 (1969).

\bibitem{Clauser+1:74}
J. F. Clauser and M. A. Horne,
\prd~\textbf{10}, 526 (1974).

\bibitem{Tabia+1:11}
G. Tabia and B.-G. Englert,
Phys.~Lett.~A~\textbf{375}, 817 (2011).

\bibitem{Shang+2:14}
J. Shang, H. K. Ng, and B.-G. Englert,
eprint arXiv:1405.5350 [quant-ph] (2014).

\bibitem{Dai+4:16}
J. Dai, Y. S. Teo, Y. L. Len, H. K. Ng, and B.-G. Englert,
in preparation (2016).

\bibitem{Mogilevtsev+2:09}
D. Mogilevtsev, J. \v{R}eh\'a\v{c}ek, and Z. Hradil,
\pra~\textbf{79}, 020101 (2009).

\bibitem{Mogilevtsev:10}
D. Mogilevtsev,
\pra~\textbf{82}, 021807 (2010).

\bibitem{Sim:15}
J. Y. Sim,
\textit{Self-calibrating Quantum State Estimation\/},
(B.Sc.\ thesis, Singapore, 2015).

\bibitem{Plante:91}
A. Plante, Canad.\ J.\ Statist.\ \textbf{19}, 389 (1991).

\bibitem{Jaynes:76}
E. T. Jaynes, \textit{Confidence Intervals vs Bayesian Intervals\/},
pp.~175--267 in
\textit{Foundations of Probability Theory, Statistical Inference, and
  Statistical Theories of Science}, vol.~II
(Reidel Publishing Company, Dordrecht, 1976),
edited by W. L. Harper and C. A. Hooker.

\bibitem{VanderPlas:14}
J. VanderPlas, eprint arXiv:1411.5018 [astro-ph.IM] (2014).

\bibitem{Chernoff-quote}
H. Chernoff as quoted on p.~178 in
\textit{The theory that would not die\/}
(Yale UP, 2011) by S. B. McGrayne.

\bibitem{Jaynes:03}
E. T. Jaynes, \textit{Probability Theory---The Logic of Science}
(Cambridge UP, 2003)

\bibitem{NJPexpert:16}
As asserted by an expert reviewer of a research journal.

\bibitem{Christandl+1:12}
M. Christandl and R. Renner, \prl\ \textbf{109}, 120403 (2012).

\bibitem{Blume-Kohout:12}
R. Blume-Kohout, eprint arXiv:1202:5270 [quant-ph] (2012).

\bibitem{note11}
For the properties of two-qubit states and their
classification, see
B.-G. Englert and N. Metwally, \textit{Kinematics of qubit pairs\/},
chapter 2 in \textit{Mathematics of Quantum Computation\/}
(Boca Raton: Chapman and Hall, 2002),
edited by G. Chen and R. K. Brylinski.

\end{thebibliography}
\end{document}